# A Solid-Based Approach for Modeling Simple Yield-Stress Fluids: Rheological Transitions, Overshoot and Relaxation

Jehyeok Choi[1], Ju Min Kim[2, 3*] and Kwang Soo Cho[1*]
[1]Department of Polymer Science and Engineering,
Kyungpook National University, Daegu, Korea
[2]Department of Chemical Engineering,
Ajou University, Suwon, Korea
[3]Department of Energy Systems Research,
Ajou University, Suwon, Korea
*Co-corresponding authors: jumin@ajou.ac.kr (Ju Min Kim),
polphy@knu.ac.kr (Kwang Soo Cho)

## Abstract

Yield-stress fluids are ubiquitous and encountered in diverse fields ranging from natural muddy flows to industrial applications such as secondary battery electrode slurries and direct ink writing. Despite the proposal of various constitutive equations, few models have been shown to successfully predict both steady and transient rheological behaviors in yield-stress fluids. In this study, a constitutive equation is hereby proposed, offering a comprehensive description of the rheological characteristics observed in simple yield-stress fluids, excluding thixotropy, such as the Carbopol dispersion. The constitutive equation is derived from a Zener-type viscoelastic solid element combined with an additional linear dashpot connected in parallel, together with a nonlinear viscosity model, a flow rule, an evolution equation for the back stress, and the Kröner–Lee decomposition. This combination satisfies the principle of material frame invariance. The proposed model successfully reproduces the rheological characteristics qualitatively in a manner consistent with experimental observations conducted during start-up shear, creep, and stress relaxation tests. In particular, the present viscoelastic solid-based constitutive equation is shown to accurately predict stress overshoot during start-up shear. Importantly, the overshoot is found to originate from a homogeneous mechanism in which normal stress difference enhances the stress invariant and thereby accelerates the plastic response, rather than from isotropic hardening or spatially heterogeneous microstructural evolution. This study is expected to facilitate a deeper understanding of the intricate dynamics governing the flow of yield-stress fluids.

# 1. Introduction

Yielding behavior is widely observed in soft matter and complex fluid systems. Microgel-based materials, such as Carbopol dispersions, as well as electrode slurries used in lithium-ion batteries, are classified as yield-stress fluids because they exhibit Herschel-Bulkley behavior under steady shear flow [Zhao *et al.*, 2022; Kim *et al.*, 2022]. However, the rheological behavior of these materials cannot be fully elucidated based solely on steady-state properties such as shear viscosity, and their responses under unsteady conditions involve complex physical mechanisms.

For instance, the stress of the Carbopol dispersion exhibits an overshoot in the start-up shear flow, depending on the flow conditions [Dinkgreve *et al.*, 2016; Dinkgreve *et al.*, 2018; Pagani *et al.*, 2024], although the steady-state stress attained after the start-up dynamics follows the Herschel–Bulkley relation with respect to the applied shear rate [Dinkgreve, Denn and Bonn, 2017]. In a stress relaxation experiment conducted subsequent to the cessation of the start-up shear test, the stress did not fully diminish to zero; instead, a finite, nonzero, fully relaxed stress was observed [Vinutha *et al.*, 2024]. Furthermore, in a creep test, a transition from solid-like behavior, characterized by strain saturation to a finite value, to fluid-like behavior, indicated by a continuous increase in strain over time occurred, depending on the magnitude of the applied stress [Ketz, Prud'homme and Graessley, 1988]. These observations suggest that the mechanical behavior of yield-stress fluids cannot be captured easily using a single steady-state relation, such as the Herschel–Bulkley model.

Consequently, numerous studies have been conducted to develop constitutive equations to accurately predict the dynamic rheological characteristics of yield-stress fluids [Saramito, 2007; Dimitriou, Ewoldt and McKinley, 2013; Dimitriou and McKinley, 2019; Kamani *et al.*, 2021; Griebler *et al.*, 2025]. These studies effectively reproduced the Herschel–Bulkley behavior in steady shear, transient dynamics [Saramito, 2007; Dimitriou, Ewoldt, and McKinley, 2013; Dimitriou and McKinley, 2019] and the overshoot of the loss modulus in large oscillatory shear flows [Kamani *et al.*, 2021; Griebler *et al.*, 2025] in yield-stress fluids. However, it is imperative to acknowledge that these models are not universally applicable in predicting all the outcomes of creep or stress relaxation experiments precisely. For instance, although kinematic hardening (KH)-based elastoviscoplastic models invoke isotropic hardening/softening (thixotropy), which involves numerous parameters, to reproduce stress overshoots in start-up shear [Dimitriou and McKinley, 2014], the stress overshoot was observed in the start-up shear of simple yield-stress fluids with weakly or negligible thixotropy, such as Carbopol dispersion [Dinkgreve *et al.*, 2016; Dinkgreve *et al.*, 2018; Pagani *et al.*, 2024]. This finding provides support for the hypothesis that, in the case of simple yield-stress fluids, transient overshoots must not be attributed exclusively to thixotropic evolution. Furthermore, in stress relaxation experiments, the stress ceases to decrease and freezes once the unyielded state is reached [Saramito, 2007], or stress relaxation does not occur [Kamani *et al.*, 2021]. This prediction is incongruent with the finite, non-zero, and fully relaxed stresses observed in recent experiments with actual yield-stress fluids [Vinutha *et al.*, 2024]. These considerations provide the impetus for developing a concise but unified constitutive framework for the simple yield-stress fluids to accurately predict the actual experimental observations in the creep and stress relaxation tests.

Yield-stress fluids such as the Carbopol dispersion have long been the subject of scientific debate regarding the boundaries between fluids and solids. For instance, should the pre-yield state be regarded as a solid or a fluid? [Barnes, 1999; Bonn and Denn, 2009]. Conversely, the categorization of a yield-stress fluid as a discrete category of fluid following yielding is a matter of consequence. In creep tests, when the applied stress is sufficiently high, the Carbopol dispersion exhibits the fluid-like behavior of continuously increasing strain without saturation. However, in stress relaxation tests, the Carbopol dispersion exhibits nonzero fully relaxed stress, which is characteristic of solids [Vinutha *et al.*, 2024]. Therefore, from the perspective of fully relaxed stress, the Carbopol dispersion should be considered as a solid. A similar problem was encountered when considering metal plasticity. Metals exhibit nonzero fully relaxed stress in relaxation tests, and the creep test of metals shows a continuously increasing strain, similar to a fluid, if the stress exceeds the yield stress. Interestingly, in the field of metal plasticity, metals are considered solids that flow like fluids above the yield condition. The similarity in the mechanical behaviors of metals and yield-stress fluids suggests the appropriateness of modeling *yield-stress fluids* from viscoelastic solids.

To comprehend the correlation between microstructural alterations and rheological behavior when stress is applied to yield-stress fluids, the Carbopol dispersion has been used as a model yield-stress fluid. The Carbopol dispersion is a microgel-based system prepared by hydrating and neutralizing crosslinked poly(acrylic acid) in de-ionized water. The hydrated microgel particles are packed at a high concentration to form a jammed state [Jaworski *et al.*, 2022]. Considering this microstructure, the deformation of the Carbopol dispersion can be decomposed into deformation within individual microgel particles and alterations in the configuration of aggregated microgel particles. When the applied stress is low, the deformation within the particles is dominant, thus allowing for the reversible recovery of the deformation. However, when the applied stress is sufficiently high, irreversible deformation occurs because of microgel rearrangement. We define the former as the elastic strain and the latter as the plastic strain. Microstructural changes in the Carbopol dispersion are analogous to dislocation phenomena occurring along grain boundaries in metallic crystals. Additionally, the Carbopol dispersion is a hydrated microgel system, and the imposed deformation inevitably involves the viscous resistance of the interstitial solvent. Because the solvent undergoes the same macroscopic deformation as the jammed microgel system, it provides an additional dissipative pathway that operates in parallel with the gel-phase response.

This study proposes a constitutive equation that predicts the rheological behavior of a simple yield-stress fluid based on a viscoelastic solid framework with an explicit solvent-like dissipative contribution, as reflected in the microstructural characteristics of the aforementioned Carbopol dispersion. The gel-phase response can be described using the Zener model, which comprises a Kelvin-Voigt element and a spring connected in series. The deformation is decomposed into elastic and plastic strains. Because the Zener model represents the jammed microgel system discussed above and the behavior of the Carbopol dispersion must include solvent-induced changes, a linear dashpot element is connected in parallel to the Zener model to account for the viscous dissipation of the interstitial solvent, which differs from metal plasticity. In this parallel arrangement, the strain rate of the linear dashpot is the same as the macroscopic strain rate and thus contributes an additional stress proportional to the total strain rate. Accordingly, the total stress is expressed as the sum of the gel-phase stress, which is governed by the elastic strain, and the solvent stress, which is governed by the total strain rate. Additionally, the flow rule and back stress evolution equation that govern the plastic deformation are introduced. First, a one-dimensional (1D) constitutive equation is derived under simple shear. Subsequently, analytical and numerical investigations are conducted to assess the response under start-up shear, stress relaxation, and creep conditions. These analyses are performed to compare the characteristics predicted by the abovementioned model with actual experimental results.

Notably, the present 1D formulation is a scalar constitutive model customized for simple shear, in which the underlying mechanisms are represented using an additive strain decomposition. Accordingly, although the 1D model is useful for conducting analytical investigations and clarifying key physical ingredients, it is not intended to directly represent general deformation modes involving finite deformations. Hence, we extend the model to three dimensions using the multiplicative decomposition, thus resulting in a formulation consistent with the principle of material frame invariance.

In the following sections, the 1D model is extended to a general three-dimensional (3D) constitutive equation. In the context of finite deformation, the Kröner–Lee decomposition is employed instead of the additive decomposition of strain. This approach ensures that the resulting constitutive equations are consistent with the principle of material frame invariance. Consequently, a theoretical framework that is expected to be applicable to general large deformation cases is developed.

The goal of this study is to develop a constitutive equation that ultimately satisfies the following four rheological characteristics of the yield-stress fluid experimentally observed in the Carbopol dispersion:

[i] Steady state stress after stress overshoot in start-up shear tests [Dinkgreve *et al.*, 2016; Dinkgreve *et al.*, 2018; Pagani *et al.*, 2024].

[ii] Behavior similar to the Herschel-Bulkley model in steady shear tests [Dinkgreve, Denn and Bonn, 2017];

[iii] A nonzero fully relaxed stress that depends on the shear rate in a stress relaxation experiment after the start-up shear test is halted [Vinutha *et al.*, 2024];

[iv] Transition from a linear viscoelastic solid to a viscoelastic fluid in creep tests at various stress amplitudes [Ketz, Prud'homme and Graessley, 1988];

# 2. 1D Constitutive Equation

## 2.1. Outline of Modeling

The 1D constitutive equation is considered first because it is mathematically simple, thereby enabling an analytical investigation into the constitutive equation, and because the physical mechanism can be readily formulated as explicit equations. The 1D constitutive equation can be readily extended to encompass more sophisticated 3D models that can predict the general deformation and flow. To motivate the mechanical construction, Fig. 1(a) schematically illustrates the microstructure of Carbopol as a jammed assembly of swollen microgel particles (crosslinked poly(acrylic acid), PAA) with interstitial solvent occupying the spaces between particles. Guided by this physical picture, we adopt a viscoelastic solid foundation with an explicit solvent-like dissipative pathway, summarized in Fig. 1(b). Specifically, the gel response is described by a Zener element, wherein a linear spring (Spring 1) is connected in series with a Kelvin–Voigt element (Spring 2 and Dashpot 2), while an additional linear dashpot (Dashpot 1) is connected in parallel to represent viscous dissipation through the interstitial solvent.

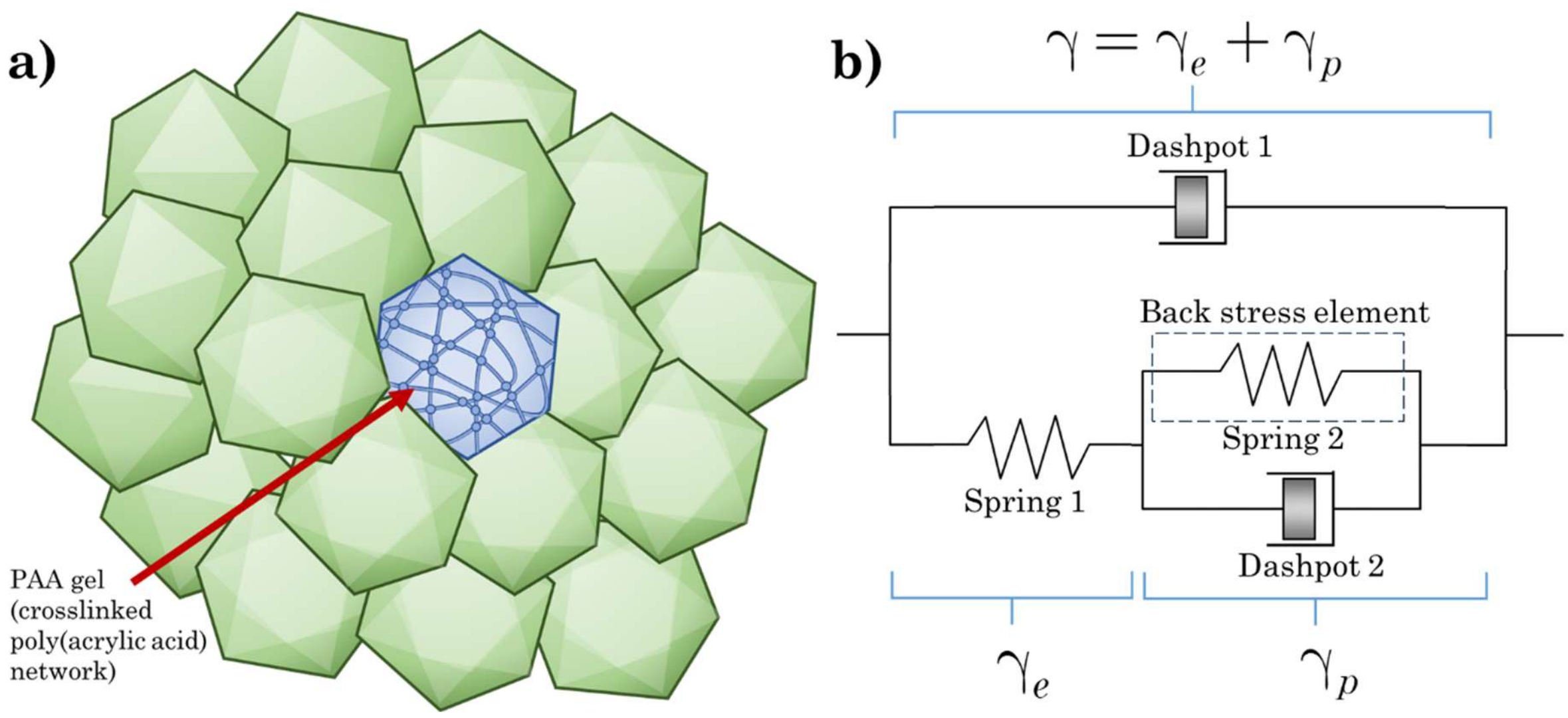

**Fig. 1.** (a) Carbopol microstructure schematic: jammed swollen PAA microgels immersed in interstitial solvent. (b) 1D mechanical analogue: Zener-type viscoelastic solid in parallel with a solvent dashpot.

The strain of Spring 1 is selected as the elastic strain $\gamma_e$, whereas that of the Kelvin-Voigt model is regarded as the plastic strain $\gamma_p$. The sum of these strains is the total strain:

$$\gamma = \gamma_e + \gamma_p. \tag{1}$$

The equation above represents the additive decomposition of strain, which is commonly used in the theory of small-deformation elasto-plasticity [Haupt, 2000; Gurtin, Fried and Anand, 2010]. Later, we will replace this additive decomposition with a multiplicative decomposition, also known as the Kröner-Lee decomposition [Kröner, 1960; Lee, 1969], in the context of a 3D constitutive equation. This approach is underpinned by the consideration of finite deformation. Equation (1) provides the evolution equation of elastic strain as follows:

$$\frac{d\gamma_e}{dt} = \frac{d\gamma}{dt} - \frac{d\gamma_p}{dt}. \tag{2}$$

Notably, $d\gamma/dt$ is controllable but $d\gamma_p/dt$ will be modeled later. The rationale behind the emphasis on the evolution equation of elastic strain is to enable extension to three dimensions, with reference to the seminal contributions of Leonov [Leonov, 1976] and the majority of models of finite-strain plasticity or viscoplasticity [Arruda and Boyce, 1993; Anand *et al.*, 2009; Dimitriou and McKinley, 2019]. To model a yield-stress fluid based on the Zener model, equations for the spring and plastic strain rate are required.

In the present model, the total shear stress $\tau$ is expressed as

$$\tau = \tau_{\mathrm{gel}} + \tau_{\mathrm{sol}}, \quad \tau_{\mathrm{gel}} \equiv G\gamma_e, \quad \tau_{\mathrm{sol}} = \eta_s \frac{d\gamma}{dt}, \tag{3}$$

where $\tau_{\mathrm{gel}}$ represents the gel phase stress; $\tau_{\mathrm{sol}}$ represents the solvent stress; $G$ is the shear modulus of Spring 1, which is connected in series with the Kelvin-Voigt element composed of Spring 2 and Dashpot 2; and $\eta_s$ is the viscosity of Dashpot 1. Because the plastic strain rate is exerted on both the spring (Spring 2) and the dashpot of the Kelvin-Voigt model, a generic flow rule is specified as follows [Tervoort, Klompen and Govaert, 1996]:

$$\frac{d\gamma_p}{dt} = \frac{\tau_{\mathrm{gel}} - \tau_B}{\eta\left(\left|\tau_{\mathrm{gel}} - \tau_B\right| / \tau_{\mathrm{o}}\right)}, \tag{4}$$

where $\eta\left(\left|\tau_{\mathrm{gel}} - \tau_B\right| / \tau_{\mathrm{o}}\right) > 0$ is a material function (viscosity) with the dimension of viscosity; $\tau_B$ is the back stress, which is the stress of Spring 2; and $\tau_{\mathrm{o}} > 0$ is a material parameter with the dimension of stress [Tervoort, Klompen and Govaert, 1996]. In accordance with the adopted Zener element, $\tau_{\mathrm{gel}} - \tau_B$ denotes the stress applied to the dashpot of the Kelvin-Voigt element.

The model can describe yield phenomena, provided that the viscosity is relatively constant, such as $\eta_{\mathrm{o}}$ when $\left|\tau_{\mathrm{gel}} - \tau_B\right| / \tau_{\mathrm{o}} < 1$, and becomes extremely small as $\left|\tau_{\mathrm{gel}} - \tau_B\right| / \tau_{\mathrm{o}}$ exceeds 1. Clearly, the dashpot of the Kelvin-Voigt element is permitted to deform substantially, which implies that the same applies to Spring 2. This suggests the necessity for a back stress model that cannot be expressed as a linear function of plastic strain.

The phenomenon of back stress can be traced to the elasto-plasticity of kinematic hardening, which is a process integral to the description of the Bauschinger effect in the context of metal plasticity [Haupt, 2000]. Here, we adopt the Armstrong-Frederick theory [Armstrong and Frederick, 1966] for the evolution equation of the back stress as follows:

$$\frac{d\tau_B}{dt} = G_B \frac{d\gamma_p}{dt} - \left|\frac{d\gamma_p}{dt}\right| \frac{\tau_B}{\gamma_B}, \tag{5}$$

where $G_B > 0$ denotes the modulus of Spring 2; $\gamma_B$ is a positive dimensionless number referred to as back strain; and $\left|d\gamma_p/dt\right|$ is employed in lieu of $d\gamma_p/dt$ because of its consistency with the 3D Armstrong-Frederick equation and the realistic creep recovery behavior. As for $d\gamma_p/dt > 0$, Eq. (5) yields:

$$\tau_B = G_B \gamma_B \left(1 - e^{-\gamma_p/\gamma_B}\right). \tag{6}$$

Equation (6), which replaces $\gamma_p$ with $\gamma$, was adopted by Cho and Kim (2003) to predict the fully relaxed stress of polymer glasses, which was based on a model comprising one elastic and one Maxwell element in parallel. Because the core of the 1D and 3D models is to model the viscosity, the relevant explanations shall be provided in the following subsection.

## 2.2. Viscosity Models

To complete the constitutive equation, the model of the viscosity $\eta\left(\left|\tau_{\text{gel}} - \tau_B\right|/\tau_o\right)$ should be provided. In this study, we adopted the Eyring viscosity model [Eyring, 1936], which is expressed as follows [Tervoort, Klompen and Govaert, 1996]:

$$\eta_{\text{Eyr},\tau}\left(\frac{\tau}{\tau_o}\right) = \eta_o \frac{|\tau|/\tau_o}{\sinh\left(|\tau|/\tau_o\right)}, \tag{7}$$

where $\eta_o$ is the zero-shear viscosity and the subscript $\tau$ indicates the viscosity model as a function of stress.

Although the Eyring viscosity model was originally introduced for simple liquids, it can be applied to the Carbopol dispersion. In a system with a high density of liquid molecules, the long-range movement of each molecule is limited by the surrounding molecules. Eyring proposed that liquid molecules can overcome this restriction by overcoming a certain energy barrier [Eyring, 1936]. Similarly, within the Carbopol dispersion, the configuration of aggregated microgels varies exclusively when the stress surpasses the interactions between the microgels. Consequently, the Eyring viscosity model is expected to incorporate the yield behavior into the constitutive equation.

However, the Eyring viscosity model is originally a function of stress and can be converted to a function of shear rate via the relation $\tau = \eta\dot{\gamma}$. Thus,

$$\tau = \tau_o \sinh^{-1}\left(\eta_o\dot{\gamma}/\tau_o\right) = \eta_o\left[\frac{\tau_o}{\eta_o\dot{\gamma}}\sinh^{-1}\left(\frac{\eta_o\dot{\gamma}}{\tau_o}\right)\right]\dot{\gamma}. \tag{8}$$

Therefore, the Eyring viscosity model as a function of shear rate is expressed as

$$\eta_{\mathrm{Eyr},\dot{\gamma}} = \eta_o \left[ \frac{\tau_o}{\eta_o \dot{\gamma}} \sinh^{-1}\left( \frac{\eta_o \dot{\gamma}}{\tau_o} \right) \right] \approx \frac{\eta_o}{\left[ 1 + \left( \eta_o \dot{\gamma} / 2.672 \tau_o \right)^{1.302} \right]^{0.6272}}, \tag{9}$$

where the subscript $\dot{\gamma}$ indicates the viscosity model as a function of shear rate. The approximation shown in Eq. (9) is a particular case of the Carreau-Yasuda model with shear-thinning indices of 1.302 and 0.6272. Concomitantly, considering the adjustable nature of the shear thinning indices within the Carreau-Yasuda model, one should convert the model into a function of stress. This modification is necessary to enhance the flexibility of the constitutive equation. Subsequently, we will demonstrate that the parameters of the viscosity model can be determined from the steady stress data of the start-up shear test.

The Carreau-Yasuda model is preferred over the Eyring model in the rheology community because the former is more flexible for fitting experimental data. The Carreau-Yasuda viscosity model is expressed as

$$\eta_{\mathrm{CY},\dot{\gamma}}\left( \dot{\gamma} \right) = \frac{\eta_o}{\left[ 1 + \left( \lambda \dot{\gamma} \right)^{\alpha} \right]^{\beta}}, \tag{10}$$

where $\eta_o$ is the zero-shear viscosity; $\dot{\gamma}$ is the shear rate; $\lambda$ is the characteristic time; and the exponents $\alpha$ and $\beta$ obey

$$\alpha > 0; \quad \beta > 0; \quad \alpha\beta \equiv n < 1. \tag{11}$$

Compared with the Eyring model, the Carreau-Yasuda model shows that $\lambda \sim \eta_o / \tau_o$. To adopt the Carreau-Yasuda model in our formalism, we must express it as a function of stress.

The shear stress $\tau$ is expressed as

$$\tau = \eta_{\mathrm{CY},\dot{\gamma}} \dot{\gamma} = \frac{\eta_o \dot{\gamma}}{\left[ 1 + \left( \eta_o \dot{\gamma} / \tau_o \right)^{\alpha} \right]^{\beta}}. \tag{12}$$

Here, we define the characteristic time as $\lambda \equiv \eta_o / \tau_o$. As shown in Appendix A, Eq. (10) can be approximated as follows:

$$\eta_{\mathrm{CY},\tau}(\tau)=\frac{\eta_{\mathrm{o}}}{\left[1+(\tau/\tau_{\mathrm{o}})^{\mu}\right]^{\nu}}, \tag{13}$$

where the exponents $\mu$ and $\nu$ are adjustable parameters that satisfy Eq. (A6).

The two models expressed in Eqs. (7) and (13) predict $\eta \approx \eta_{\mathrm{o}}$ in the low stress regime ($\tau \ll \tau_{\mathrm{o}}$) while the viscosity decreases rapidly as the stress exceeds $\tau_{\mathrm{o}}$. Therefore, the two viscosity models are consistent with our constitutive equation. Thus, one may assume that $\tau_{\mathrm{o}}$ is the yield stress $\tau_y$ defined in the Herschel-Bulkley plot of steady state stress. This will be verified in Section 3.

The emergence of the first Newtonian plateau can be interpreted as a consequence of the predominant binding interaction between microgels, thus resulting in the collective motion of their clusters ($\tau < \tau_{\mathrm{o}}$). As stress increases, these clusters disintegrate, thus causing a rapid decrease in viscosity ($\tau \sim \tau_{\mathrm{o}}$). Meanwhile, as stress increases further, the clusters of microgel particles are fully destroyed into individual microgel particles and the motion of individual microgel particles triggers the emergence of a second Newtonian plateau ($\tau \gg \tau_{\mathrm{o}}$). Barnes (1999) discovered that the apparent viscosity of the Carbopol dispersion comprises two Newtonian plateaus. In the present model, the second Newtonian plateau can emerge from Dashpot 1.

## 2.3. Summary of Constitutive Equation

The constitutive equation proposed in this study includes the following set of differential equations:

$$\begin{aligned}
&\tau=\tau_{\mathrm{gel}}+\tau_{\mathrm{sol}}; \quad \tau_{\mathrm{gel}}=G\gamma_e; \quad \tau_{\mathrm{sol}}=\eta_s\frac{d\gamma}{dt}; \quad \frac{d\gamma_e}{dt}=\frac{d\gamma}{dt}-\frac{d\gamma_p}{dt}; \\
&\frac{d\tau_B}{dt}=G_B\frac{d\gamma_p}{dt}-\left|\frac{d\gamma_p}{dt}\right|\frac{\tau_B}{\gamma_B}; \quad \frac{d\gamma_p}{dt}=\frac{\tau_{\mathrm{gel}}-\tau_B}{\eta\left(\left|\tau_{\mathrm{gel}}-\tau_B\right|/\tau_{\mathrm{o}}\right)}.
\end{aligned} \tag{14}$$

We will test this constitutive equation for the following two viscosity models:

$$\begin{aligned}
&\eta_{\mathrm{Eyr}}=\eta_{\mathrm{o}}\frac{\xi}{\sinh\xi} \quad \text{Eyring viscosity} \\
&\eta_{\mathrm{CY}}=\frac{\eta_{\mathrm{o}}}{\left(1+\xi^{\mu}\right)^{\nu}} \quad \text{Carreau-Yasuda viscosity}
\end{aligned}, \quad \xi \equiv \frac{\left|\tau_{\mathrm{gel}}-\tau_B\right|}{\tau_{\mathrm{o}}}. \tag{15}$$

# 3. Test of Constitutive Equation

## 3.1. Steady State Stress in Start-up Shear Test

Consider a start-up shear test whose strain is expressed as $\gamma(t)=\dot{\gamma}_o t\Theta(t)$, where $\dot{\gamma}_o>0$ is a constant and $\Theta(t)$ is the unit step function. In the start-up shear test, stress reaches a constant steady-state value over a long duration. Because stress is a function of elastic strain, the elastic strain remains constant at steady state. The observed experimental data [Dinkgreve *et al.*, 2016; Dinkgreve *et al.*, 2018; Pagani *et al.*, 2024] can be summarized as follows:

$$\frac{d\tau}{dt}=0 \quad \Rightarrow \quad \frac{d\gamma_e}{dt}=0 \text{ for } \quad t\to\infty. \tag{16}$$

Notably, $\tau_{sol}$ is constant for the start-up shear test. In steady state, $d\gamma_e/dt=0$; thus, $d\gamma/dt=d\gamma_p/dt=\dot{\gamma}_o$. Therefore, Eqs. (14)$_4$ and (14)$_6$ yield

$$\frac{\tau_{gel}-\tau_B}{\eta\left[\left(\tau_{gel}-\tau_B\right)/\tau_o\right]}=\dot{\gamma}_o. \tag{17}$$

Here, we used $\tau_{gel}>\tau_B$ in the start-up shear test.

If the Eyring viscosity is adopted then Eq. (17) becomes

$$\frac{\tau_o}{\eta_o}\sinh\frac{\tau_{gel}-\tau_B}{\tau_o}=\dot{\gamma}_o. \tag{18}$$

Because the hyperbolic sine function is a monotonically increasing function, for a specified value of $\dot{\gamma}_o>0$, the value of $\tau_{gel}-\tau_B$ that satisfies $d\gamma_e/dt=0$ is uniquely determined. If the unique solution is denoted by $\tau_{gel}^*-\tau_B^*$ then we have

$$\begin{aligned}\tau(\infty)&=\tau_{gel}^*+\tau_{sol}=\tau_B^*+\tau_o\sinh^{-1}\frac{\eta_o\dot{\gamma}_o}{\tau_o}+\eta_s\dot{\gamma}_o\\ &\approx\tau_B^*+\frac{\eta_o\dot{\gamma}_o}{\left[1+\left(\eta_o\dot{\gamma}_o/2.672\tau_o\right)^{1.302}\right]^{0.6272}}+\eta_s\dot{\gamma}_o,\end{aligned} \tag{19}$$

where $\sinh^{-1}(x)$ is the inverse function of the hyperbolic sine. Notably, Eq. (19) is similar to the equation for the Herschel-Bulkley model.

The Herschel-Bulkley model represents the relation between the imposed shear rate $\dot{\gamma}_o$ in a steady shear test and the corresponding steady state stress $\tau(\infty)$ as follows:

$$\tau(\infty) = \tau_y + k\dot{\gamma}_o^m, \tag{20}$$

where $\tau_y > 0$ is the yield stress; $k$ is the consistency index; and the exponent $m$ is the power-law index. The fluid is shear-thinning for $m < 1$, a Bingham fluid with a constant shear viscosity for $m = 1$ and $\tau_y > 0$, and shear-thickening for $m > 1$. In many yield-stress fluids, the exponent $m$ is less than unity.

The $\sinh^{-1}(x)$ term in Eq. (19) becomes $\eta_o\dot{\gamma}_o$ when the shear rate is low, whereas it is proportional to $\dot{\gamma}_o^{0.1834}$ when the shear rate is high. Notably, $\tau_B^* \gg (\eta_o + \eta_s)\dot{\gamma}_o$ at low shear rates and the $\eta_s\dot{\gamma}_o$ term dominates at high shear rates. Therefore, by comparing Eqs. (19) and (20), the yield stress $\tau_y$ corresponds to $\tau_B^*$ and $k\dot{\gamma}_o^m \approx \eta_s\dot{\gamma}_o$.

Because $\tau_{\text{gel}}^* - \tau_B^*$ is uniquely determined for a specified value of $\dot{\gamma}_o$, the two stress values are those at $t \to \infty$: $\tau_{\text{gel}}^* = \tau_{\text{gel}}(\infty)$ and $\tau_B^* = \tau_B(\infty)$. Based on Eq. (6), when the plastic strain becomes sufficiently large after a long duration, the back stress in the steady state becomes $\tau_B(\infty) = G_B\gamma_B$.

In the case of the Carreau-Yasuda viscosity, the condition $d\gamma_e/dt = 0$ yields

$$\frac{\tau_{\text{gel}} - \tau_B}{\eta_{\text{CY}}\left[\left(\tau_{\text{gel}} - \tau_B\right)/\tau_o\right]} = \dot{\gamma}_o. \tag{21}$$

Because $\eta_{\text{CY}}$ is a decreasing function of stress, the left-hand side of Eq. (21) is an increasing function of stress; thus the value of $\tau_{\text{gel}} - \tau_B$ satisfying $d\gamma_e/dt = 0$ is uniquely determined. Therefore, the unique solution is as follows:

$$\begin{aligned}\tau(\infty) &= \tau_{\text{gel}}^* + \tau_{\text{sol}} = \tau_B^* + \eta_{\text{CY}}\left[\left(\tau_{\text{gel}} - \tau_B\right)/\tau_o\right]\dot{\gamma}_o + \eta_s\dot{\gamma}_o \\ &= \tau_B^* + \eta_{\text{CY},\dot{\gamma}}\left(\dot{\gamma}_o\right)\dot{\gamma}_o + \eta_s\dot{\gamma}_o = \tau_B^* + \frac{\eta_o\dot{\gamma}_o}{\left[1 + \left(\eta_o\dot{\gamma}_o/\tau_o\right)^\alpha\right]^\beta} + \eta_s\dot{\gamma}_o.\end{aligned} \tag{22}$$

As in the Eyring viscosity model, the two stress values are those at $t \to \infty$: $\tau_{\text{gel}}^* = \tau_{\text{gel}}(\infty)$ and $\tau_B^* = \tau_B(\infty)$. Notably, Eq. (22) is similar to the Herschel-Bulkley model. Equation (22) becomes $\tau(\infty) \approx \tau_B^*$ at zero shear rate, whereas it becomes $\tau(\infty) \approx \eta_s \dot{\gamma}_o$ at extremely high shear rates.

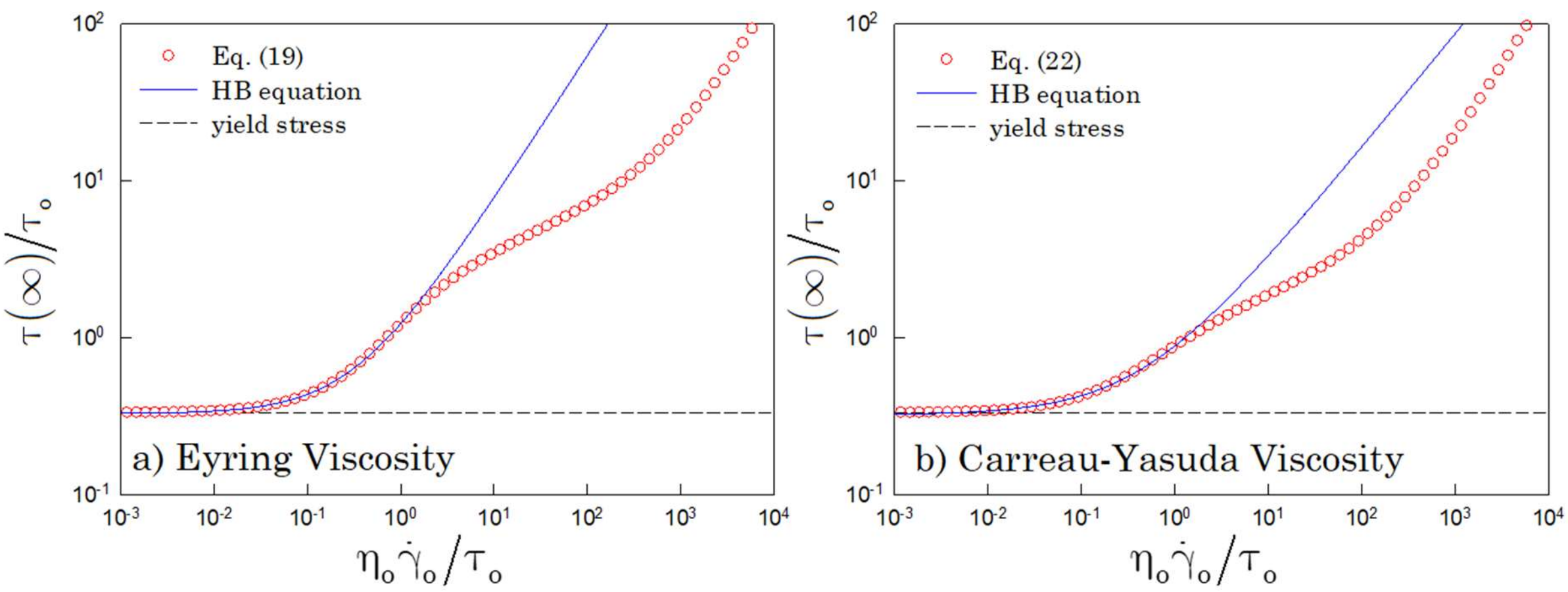

**Fig. 2**. Steady state stresses calculated from (a) Eyring viscosity and (b) Carreau-Yasuda viscosity plotted as a function of normalized shear rate $\eta_o \dot{\gamma}_o / \tau_o$. Red circles represent steady state stresses obtained from Eqs. (19) and (22), blue lines represent the best fit of the Herschel-Bulkley equation for $\eta_o \dot{\gamma}_o / \tau_o < 1$, and dotted black lines represent yield stress $\tau_B(\infty)/\tau_o \left(= \tau_y / \tau_o\right)$. The parameters are set as follows: $\tau_B(\infty) = 10\,\mathrm{Pa}$, $\tau_o = 30\,\mathrm{Pa}$, $\eta_o = 70\,\mathrm{Pa \cdot s}$, $\eta_s = 1\,\mathrm{Pa \cdot s}$, $\alpha = 0.9$, and $\beta = 0.9$.

Figure 2 shows the steady state stress obtained from Eqs. (19) and (22). The blue lines represent the best fit of the Herschel-Bulkley equation for $\eta_o \dot{\gamma}_o / \tau_o < 1$. In Figs. 2(a) and 2(b), when $\eta_o \dot{\gamma}_o / \tau_o \ll 1$, the steady state stress reflects the Herschel-Bulkley equation, thus resulting in $\tau_y = \tau_B(\infty)$. When $\eta_o \dot{\gamma}_o / \tau_o > 1$, the steady state stress appears to deviate from the Herschel-Bulkley equation. For $\eta_o \dot{\gamma}_o / \tau_o > 1$, the Herschel-Bulkley equation is a power-law function, with the steady state stress exhibiting a constant slope with respect to the shear rate in a double-log plot. However, the slope of the stresses obtained from the two viscosity models changed as the shear rate increased. Notably, the curves of both models transformed from concave to convex when $\eta_o \dot{\gamma}_o / \tau_o > 10$. The change in convexity is due to the existence of $\eta_s$, a second Newtonian plateau.

Because the Herschel-Bulkley equation is an empirical model, our calculations deviated from it, which is not surprising. Therefore, experimental validation is required. Wall slip is a well-known complication in Carbopol rheometry, especially at low shear rates on smooth boundaries [Divoux, Barentin and Manneville, 2011; Bonn *et al.*, 2017]. At higher imposed deformation rates or stresses, deviations should not be attributed generically to wall slip; in Carbopol, the better documented complications are sample fracture/localized deformation and, in parallel-plate geometries, edge fracture or other free-surface instabilities [Piau, 2007; Medina-Bañuelos *et al.*, 2022]. Changes in the material may exhibit such high-shear behavior in a moderate shear range. If the fringe chains on the surface of the microgel particles are short, then the binding interaction between the particles may be reduced, and such high-shear behavior may be observed. Based on experimental data for Carbopol 940, we observed a transition in convexity, as shown in Fig. 3.

The steady shear data in Fig. 3 were obtained at 23 °C using a rotational rheometer (DHR-3, TA Instruments) featuring a parallel-plate geometry (diameter = 40 mm) for a 0.2 wt% Carbopol 940 aqueous solution. The sample was prepared by dispersing Carbopol microgels in deionized water and stirring at 150 rpm for 24 h, followed by the gradual addition of 20 wt% NaOH to adjust the pH to 7 and additional stirring for 6 h. Prior to the steady shear measurement, a pre-shear protocol was applied, and the sample was allowed to settle for 1 min. Subsequently, a descending shear rate sweep from 1000 to 0.01 /s was performed, and the steady-state stress at each step was extracted under the steady state sensing mode.

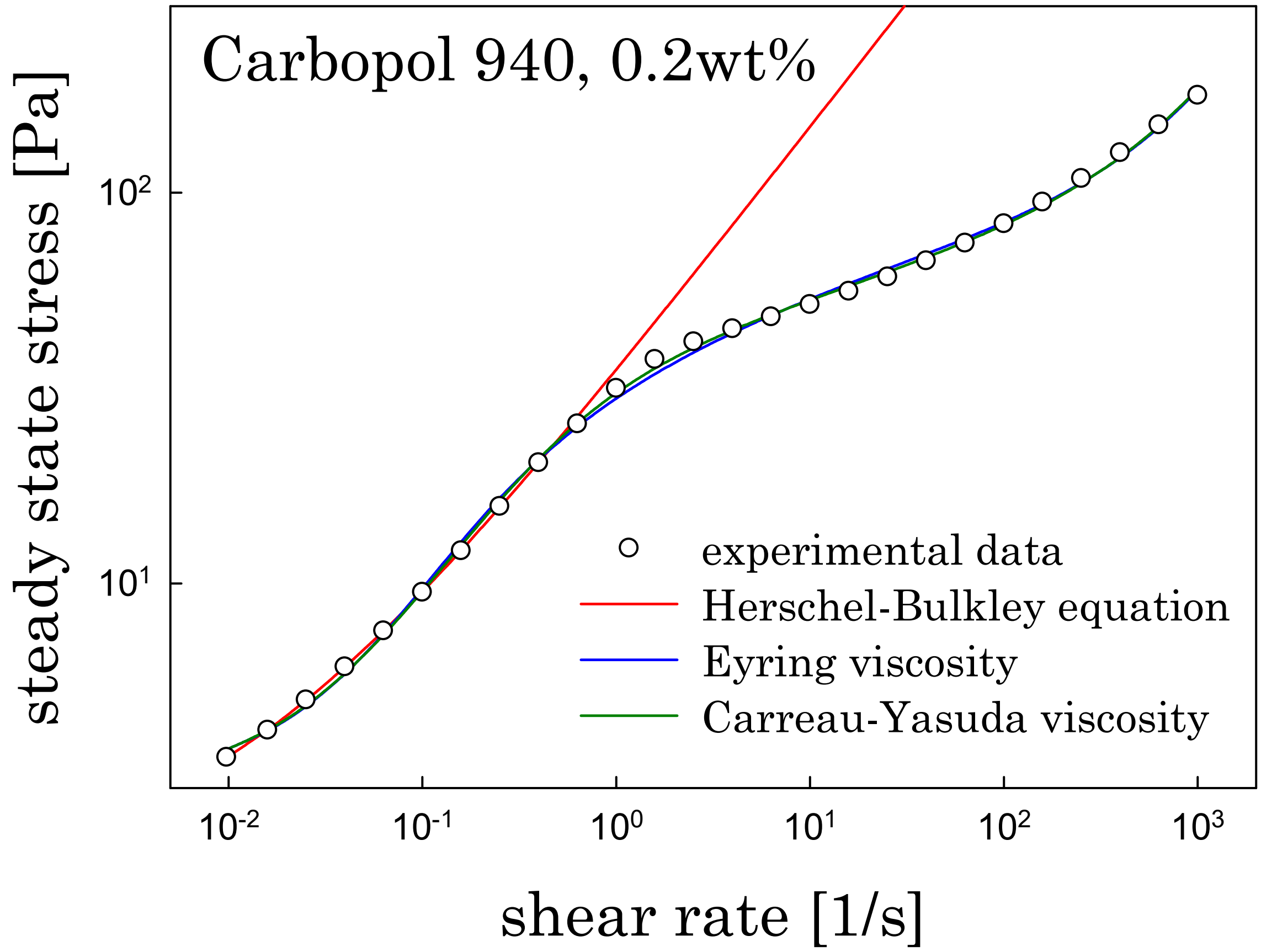

**Fig. 3.** Steady state stress of Carbopol 940 0.2 wt% solution as a function of shear rate obtained from steady shear test. Black circle represents experimental data; red line represents best fit of Herschel-Bulkley equation for $\dot{\gamma} < 1/\mathrm{s}$; and blue and green lines represent fitting results using Eyring and Carreau-Yasuda viscosities in the general Herschel-Bulkley equation, respectively.

Figure 3 shows that the parameters of the viscosity models can be determined from the plot of steady state stress as a function of shear rate. Furthermore, one may assume that Eq. (17) yields the generalized Herschel–Bulkley equation as follows:

$$\begin{aligned}\tau^{(\infty)}(\dot{\gamma}) &= \tau_B^{(\infty)} + \eta_\tau\left(\left|\tau_{\mathrm{gel}}^{(\infty)}(\dot{\gamma}) - \tau_B^{(\infty)}\right| \Big/ \tau_{\mathrm{o}}\right)\dot{\gamma} + \eta_s\dot{\gamma} \\ &= \tau_B^{(\infty)} + \eta_{\dot{\gamma}}(\dot{\gamma})\dot{\gamma} + \eta_s\dot{\gamma},\end{aligned} \tag{23}$$

where $\tau^{(\infty)}(\dot{\gamma})$ is the steady state stress at a constant shear rate $\dot{\gamma}$ and $\tau_B^{(\infty)}$ is the back stress at infinite time, which is independent of the shear rate. The parameters of the viscosity model can be obtained via a regression analysis of the steady state stress using Eq. (23).

Interestingly, the agreement between our model and the Herschel-Bulkley equation holds for $\eta_o\dot{\gamma}/\tau_o < 1$. Therefore, $\tau_o$ is not the yield stress from the perspective of the Herschel-Bulkley model; instead, it indicates the beginning of shear thinning. This indicates an abrupt increase in the plastic strain rate when the flow rule is considered.

In the 1D constitutive equation, predicting the stress overshoot using the Eyring and Carreau-Yasuda viscosity models is challenging, as shown in Fig. 4. Equation (6) implies that $\tau_B(t)$ is a monotonically increasing function of time. The appearance of stress overshoot indicates the existence of at least two solutions to Eq. (17), which is a consequence of $d\tau/dt = 0$. However, our two viscosity models do not allow for two solutions because viscosity is a monotonic function of the stress difference $\tau_{\text{gel}} - \tau_B$. However, we will show that our 3D model can predict the stress overshoot.

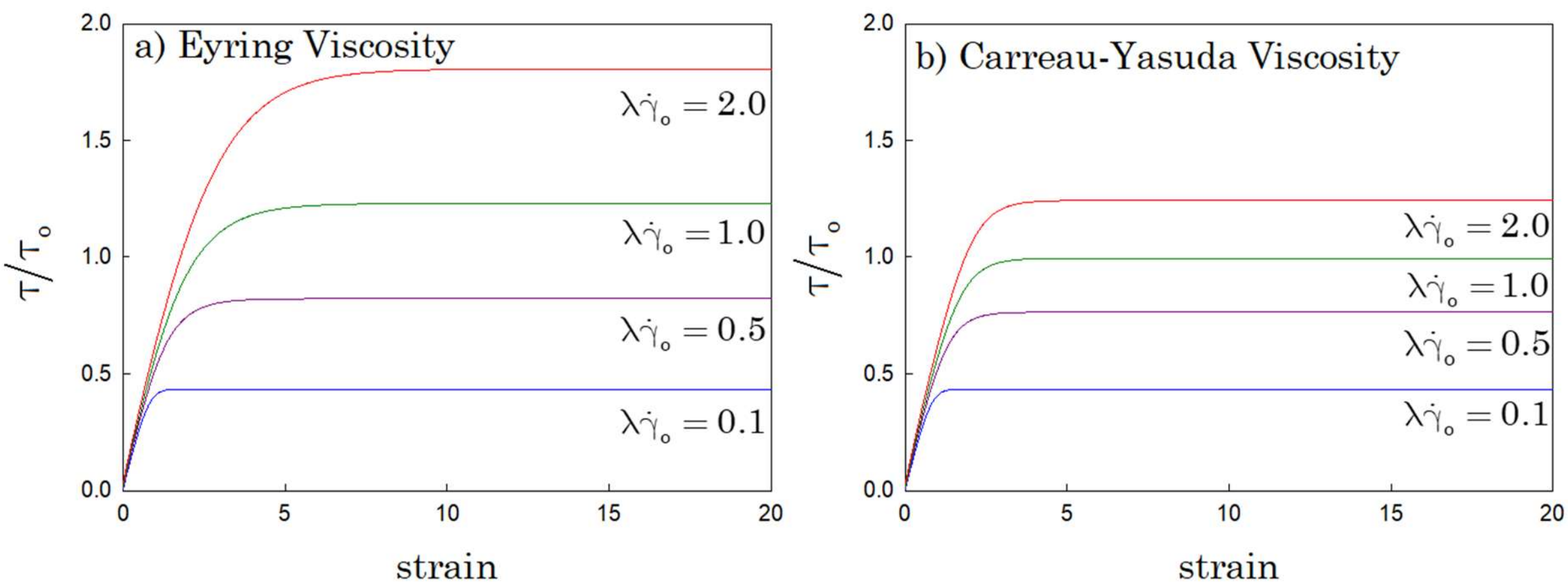

**Fig. 4.** Shear stresses of start-up shear calculated via numerical integration of Eq. (14) with (a) Eyring viscosity and (b) Carreau-Yasuda viscosity models. The parameters are set as follows: $\gamma_B = 0.1$, $G = 20\,\text{Pa}$, $G_B = 100\,\text{Pa}$, $\tau_o = 30\,\text{Pa}$, $\eta_o = 70\,\text{Pa}\cdot\text{s}$, $\eta_s = 1\,\text{Pa}\cdot\text{s}$, $\mu = 2.5$, $\nu = 1.5$ and $\lambda = \eta_o/\tau_o$.

Figure 4 shows the numerical integration results of the constitutive equation shown in Eq. (14), at a constant shear rate. As time (or strain) increased, the shear stress increased and approached a steady state value. The steady state stress $\tau(t \to \infty)$ as a function of the shear rate $\dot{\gamma}_o$ is consistent with Eqs. (19) and (22) for the Eyring and Carreau–Yasuda viscosity models, respectively. Consistent with the analytical argument above, no stress overshoot was observed for either viscosity model; instead, the stress increased monotonically toward its steady value. Additionally, the strain required to reach a steady state increased with the shear rate. By contrast, we will show that the 3D extension, which incorporates tensorial-stress evolution, can exhibit stress overshoot under start-up shear.

### 3.2. Fully Relaxed Stress of Relaxation Experiment

Consider a test in which a constant shear rate $\dot{\gamma}_o$ is applied for $0 \le t \le t_o$ (start-up shear test), after which the strain is fixed, i.e., $\gamma(t) = \gamma(t_o)$ for $t_o < t$. Stress relaxation is observed as a function of $t_r \equiv t - t_o$. For $t_r > 0$, since the total strain is fixed and the shear rate is zero, we have $d\gamma/dt = 0$. Thus, Eq. (14) is reduced to

$$
\begin{aligned}
&\frac{d\gamma_e}{dt} = \frac{1}{G}\frac{d\tau}{dt} = -\frac{d\gamma_p}{dt}; \\
&\frac{d\tau_B}{dt} = G_B \frac{d\gamma_p}{dt} - \left|\frac{d\gamma_p}{dt}\right| \frac{\tau_B}{\gamma_B}; \\
&\frac{d\gamma_p}{dt} = \frac{\tau_{\text{gel}} - \tau_B}{\eta\left(\left|\tau_{\text{gel}} - \tau_B\right| / \tau_o\right)}.
\end{aligned}
\tag{24}
$$

Here, we are interested in the non-zero fully relaxed stress which is the stress at $t \to \infty$ and $d\gamma_e/dt = 0$.

We assume that the pre-shearing time $t_o$ is sufficiently long such that the stress reaches a steady state. Subsequently, the initial stress $\tau(t = t_o)$ can be approximated using the Herschel–Bulkley model.

$$
\tau(t_o) \approx \tau_B(t_o) + k\dot{\gamma}_o^m. \tag{25}
$$

Similarly, it can be approximated using generalized Herschel–Bulkley equations [Eyring and Carreau–Yasuda models, Eq. (23)]. However, we used a simpler approximation to easily understand the effect of the pre-shear rate on the fully relaxed stress. Here, $t_o$ is the time at which steady shearing is terminated. Because the condition for full relaxation is $d\gamma_e/dt = d\gamma_p/dt = 0$, Eq. $(24)_3$ yields

$$
\tau(\infty) = \tau_B(\infty). \tag{26}
$$

Notably, $d\gamma_p/dt \geq 0$ and $\tau_{\text{sol}} = 0$ for $t > t_{\text{o}}$. Therefore, integrating Eqs. (24)1 and (24)2 yields

$$
\begin{aligned}
&\gamma_p(\infty) - \gamma_p(t_{\text{o}}) = \gamma_e(t_{\text{o}}) - \gamma_e(\infty) = \frac{\tau(t_{\text{o}}) - \tau(\infty)}{G}; \\
&\tau_B(t) = e^{-[\gamma_p(t) - \gamma_p(t_{\text{o}})]/\gamma_B} \tau_B(t_{\text{o}}) + G_B \gamma_B \left[1 - e^{-[\gamma_p(t) - \gamma_p(t_{\text{o}})]/\gamma_B}\right].
\end{aligned} \tag{27}
$$

Combining Eq. (26) using the two equations in Eq. (27) yields

$$
\tau_B(\infty) = \tau_B(t_{\text{o}}) + \left[G_B \gamma_B - \tau_B(t_{\text{o}})\right]\left[1 - e^{-[\tau(t_{\text{o}}) - \tau_B(\infty)]/G\gamma_B}\right]. \tag{28}
$$

Because $t_{\text{o}}$ is assumed to be sufficiently long, we can conclude that $\gamma_p(t_{\text{o}}) \gg \gamma_B$. The integration of Eq. (24)2 from $t = 0$ to $t = t_{\text{o}}$ yields

$$
\tau_B(t_{\text{o}}) = G_B \gamma_B \left[1 - e^{-\gamma_p(t_{\text{o}})/\gamma_B}\right] \approx G_B \gamma_B. \tag{29}
$$

Substituting Eq. (29) into Eq. (28) yields

$$
\tau_B(\infty) \approx \tau_B(t_{\text{o}}) \approx G_B \gamma_B. \tag{30}
$$

Subsequently, the following relation can be predicted using Eq. (25):

$$
\frac{\tau(\infty)}{\tau(t_{\text{o}})} \approx \frac{G_B \gamma_B}{G_B \gamma_B + k \dot{\gamma}_{\text{o}}^m} = \frac{1}{1 + \bar{k} \dot{\gamma}_{\text{o}}^m}, \tag{31}
$$

where

$$
\bar{k} = \frac{k}{G_B \gamma_B}. \tag{32}
$$

Specifically, the steady-state $\tau(t_{\text{o}})$ obeys Eqs. (19) and (22) for the start-up shear test and $\tau_{\text{sol}} = 0$ for the stress relaxation test, which predicts

$$\frac{\tau(\infty)}{\tau(t_o)} = \begin{cases} \dfrac{G_B\gamma_B}{G_B\gamma_B + \tau_o \sinh^{-1}\dfrac{\eta_o\dot{\gamma}_o}{\tau_o}} & \text{for Eyring viscosity} \\[2ex] \dfrac{G_B\gamma_B}{G_B\gamma_B + \dfrac{\eta_o\dot{\gamma}_o}{\left[1+\left(\eta_o\dot{\gamma}_o/\tau_o\right)^{\alpha}\right]^{\beta}}} & \text{for Carreau-Yasuda viscosity.} \end{cases} \tag{33}$$

Equations (31) and (33) imply that the normalized fully relaxed stress $\tau(\infty)/\tau(t_o)$ decreases when the pre-shearing rate $\dot{\gamma}_o$ increases. This trend is qualitatively consistent with the trend reported by Vinutha *et al.* (2024).

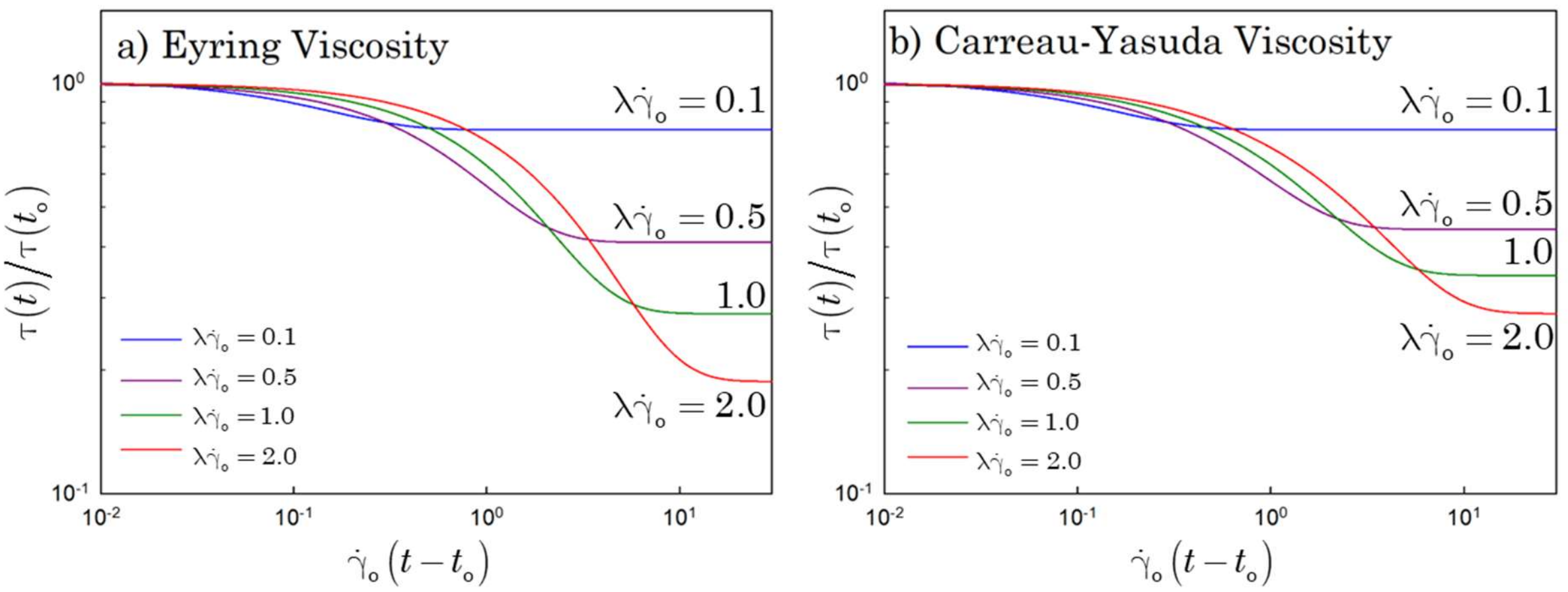

**Fig. 5.** Shear-stress relaxation as a function of dimensionless time in stress-relaxation tests obtained from (a) Eyring viscosity and (b) Carreau–Yasuda viscosity models. The parameters are set as follows: $t_o = 20/\dot{\gamma}_o$, $t_{max} = 40/\dot{\gamma}_o$, $\gamma_B = 0.1$, $G = 20\,\text{Pa}$, $G_B = 100\,\text{Pa}$, $\tau_o = 30\,\text{Pa}$, $\eta_o = 70\,\text{Pa}\cdot\text{s}$, $\eta_s = 1\,\text{Pa}\cdot\text{s}$, $\mu = 2.5$, $\nu = 1.5$ and $\lambda = \eta_o/\tau_o$.

Figure 5 shows the numerical integration results for stress relaxation. A start-up shear test was performed with a shear rate $\dot{\gamma}_o$ up to time $t = t_o$, and stress relaxation was proceeded over $t_o \le t \le t_{max}$. For both viscosity models, the fully relaxed stress was obtained using $\tau(\infty) = G_B\gamma_B$. The $\tau(\infty)/\tau(t_o)$ decreased as the pre-shearing rate $\dot{\gamma}_o$ increased. These numerical results reproduce the qualitative trend reported by Vinutha *et al.* (2024), although Fig. 5 contains model calculations only and is not a direct fit to their experimental data.

Based on the analytical and numerical analyses, our model predicts $\tau(\infty) = G_B \gamma_B$ independent of the pre-shearing rate. However, whether the same trend is observed in these experiments remains unclear. If the experimental data show the dependence of $\tau(\infty)$ on the pre-shearing rate, then the evolution equation of the back stress must be modified.

## 3.3. Transition from Solid to Fluid in Creep Test

Because stress is fixed in the creep test, we have $\tau(t) = \tau_{\text{app}}$ for $t > 0$. Thus, Eq. (14) yields

$$\frac{d\gamma}{dt} = \frac{\tau_{\text{app}} - \tau_{\text{gel}}}{\eta_s}; \quad \frac{d\gamma_e}{dt} = \frac{d\gamma}{dt} - \frac{d\gamma_p}{dt};$$
$$\frac{d\gamma_p}{dt} = \frac{\tau_{\text{gel}} - \tau_B}{\eta\left(\left|\tau_{\text{gel}} - \tau_B\right| / \tau_o\right)}; \quad \frac{d\tau_B}{dt} = \frac{d\gamma_p}{dt}\left(G_B - \frac{\tau_B}{\gamma_B}\right). \tag{34}$$

The initial conditions for the strain and back stress are as follows:

$$\gamma(0) = 0; \quad \tau_B(0) = 0. \tag{35}$$

Based on Eq. (6), to satisfy the initial conditions shown in Eq. (35)2, $\gamma_p(0) = 0$.

Strain saturation implies that the strain rate approaches zero after a sufficiently long time. Based on Eq. (34)1, $d\gamma/dt = 0$ requires that $\tau_{\text{app}} = \tau_{\text{gel}}(\infty)$. Because $\tau_{\text{gel}} = G\gamma_e$, the convergence of $\tau_{\text{gel}}$ to a constant implies that $d\gamma_e/dt = 0$. Combining this with $d\gamma/dt = 0$, Eq. (34)2 yields

$$\frac{d\gamma_p}{dt} = \frac{d\gamma}{dt} - \frac{d\gamma_e}{dt} = 0. \tag{36}$$

Notably combining Eqs. (34)3 and (36) yields $\tau_{\text{gel}}(\infty) = \tau_B(\infty) = \tau_{\text{app}}$. Because the maximum value of the back stress is $G_B\gamma_B$ (based on Eq. (6)), this constitutive equation shows solid-like creep behavior under the condition $\tau_{\text{app}} < \tau_y = G_B\gamma_B$. If the constitutive equation behaves as a solid such that the strain $\gamma_p$ saturates at a finite value, e.g., $\gamma_p(\infty) = \gamma_p^{(\max)} < \infty$, then $d\gamma_p/dt$ must be zero at $\gamma_p^{(\max)}$:

$$\tau_B(\infty) = G_B\gamma_B\left(1-e^{-\gamma_p^{(\max)}/\gamma_B}\right); \quad \gamma_p^{(\max)} = -\gamma_B \ln\left(1-\frac{\tau_{\text{app}}}{G_B\gamma_B}\right). \tag{37}$$

Notably, $\tau_{\text{app}} = \tau_{\text{gel}}(\infty)$ yields $\gamma_e(\infty) = \tau_{\text{app}}/G$. Thus, the saturated strain at $t=\infty$ is expressed as

$$\gamma(\infty) = \frac{\tau_{\text{app}}}{G} - \gamma_B \ln\left(1-\frac{\tau_{\text{app}}}{G_B\gamma_B}\right). \tag{38}$$

Here, $\gamma(\infty)$ increases with $\tau_{\text{app}}$ under the condition $\tau_{\text{app}} < G_B\gamma_B$. Therefore, the transition stress is expressed as follows:

$$\tau_{\text{tran}} \equiv G_B\gamma_B. \tag{39}$$

Here, $\tau_{\text{tran}}$ is the maximum back stress during the start-up shear test, and the back stress is the fully relaxed stress. This implies that the yield stress of the Herschel–Bulkley equation is a meaningful indicator of yield behavior because it can be regarded as the stress of the transition from a linear viscoelastic solid to a viscoelastic fluid.

Otherwise, when $\tau_{\text{app}} > \tau_y$, the saturation condition $\tau_B(\infty) = \tau_{\text{gel}}(\infty) = \tau_{\text{app}}$ cannot be satisfied. Based on Eq. (34), the evolution equation of the gel stress is expressed as follows:

$$\frac{d\tau_{\text{gel}}}{dt} = G\frac{d\gamma_e}{dt} = G\left(\frac{d\gamma}{dt} - \frac{d\gamma_p}{dt}\right) = G\left(\frac{\tau_{\text{app}}-\tau_{\text{gel}}}{\eta_s} - \frac{\tau_{\text{gel}}-\tau_B}{\eta\left(\left|\tau_{\text{gel}}-\tau_B\right|/\tau_o\right)}\right). \tag{40}$$

Because $d\gamma/dt = \left(\tau_{\text{app}}-\tau_{\text{gel}}\right)/\eta_s$ is a decreasing function of $\tau_{\text{gel}}$ while $d\gamma_p/dt = \left(\tau_{\text{gel}}-\tau_B\right)/\eta$ is an increasing function of $\tau_{\text{gel}}$, $\tau_{\text{gel}}$ converges to a value satisfying $\tau_y < \tau_{\text{gel}}(\infty) < \tau_{\text{app}}$. Therefore,

$$\left.\frac{d\gamma}{dt}\right|_{t=\infty} = \frac{\tau_{\text{app}}-\tau_{\text{gel}}(\infty)}{\eta_s} > 0 \quad \text{for } \tau_{\text{app}} > \tau_y, \tag{41}$$

which implies fluid-like creep with a non-zero steady strain rate.

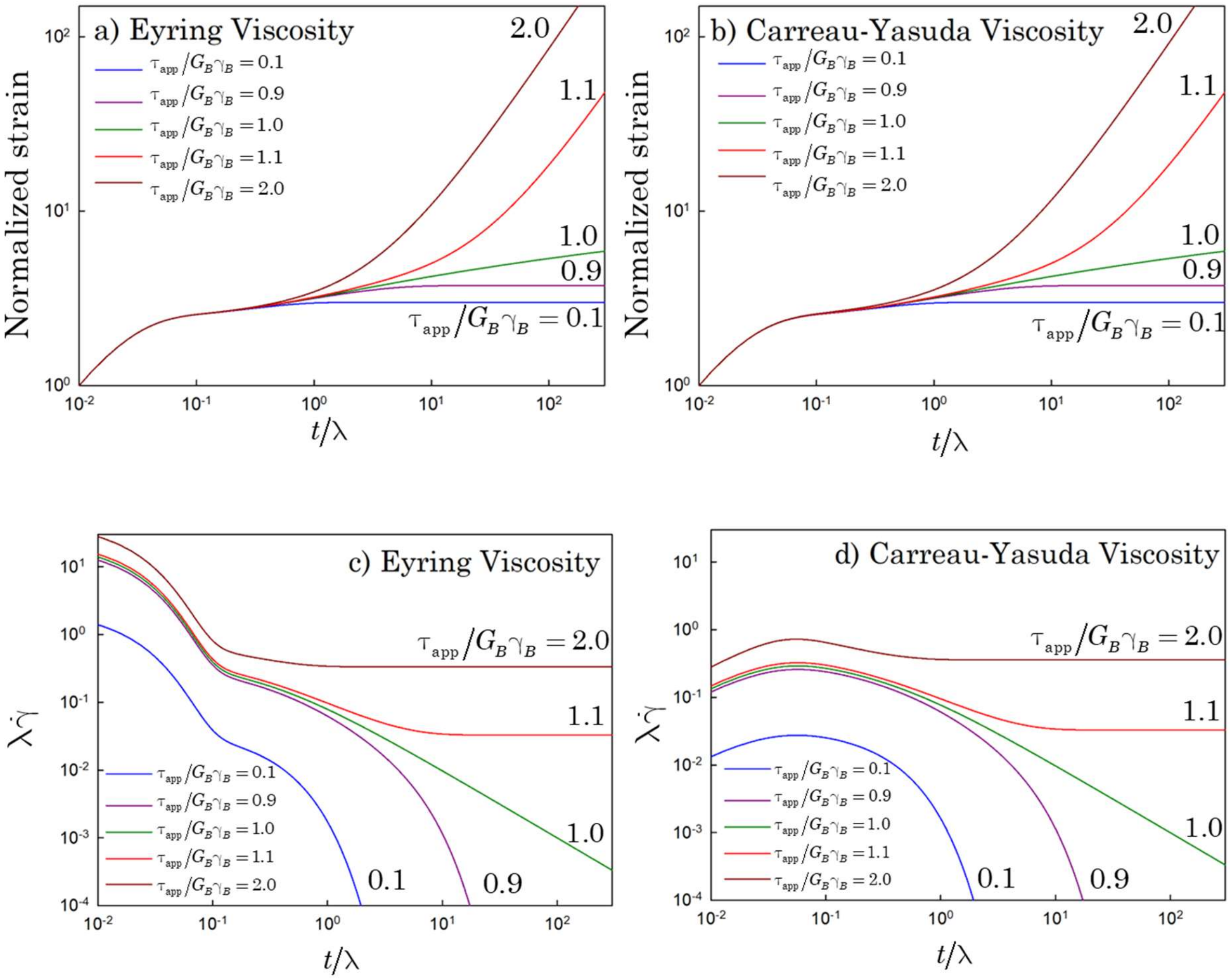

**Fig. 6**. Numerical integration results of Eq. (34) for creep test. Strains calculated using (a) Eyring viscosity and (b) Carreau–Yasuda viscosity and shear rates calculated using (c) Eyring viscosity and (d) Carreau–Yasuda viscosity. To facilitate understanding, the strains are normalized to their values at $t/\lambda = 10^{-2}$. The parameters are set as follows: $\gamma_B = 0.1$, $G = 20\,\mathrm{Pa}$, $G_B = 100\,\mathrm{Pa}$, $\tau_o = 30\,\mathrm{Pa}$, $\eta_o = 70\,\mathrm{Pa\cdot s}$, $\eta_s = 1\,\mathrm{Pa\cdot s}$, $\mu = 2.5$, $\nu = 1.5$ and $\lambda = \eta_o/\tau_o$.

Figure 6 shows the numerical results of creep transition. As shown in Figs. 6(a) and 6(b), when the applied stress $\tau_{app}$ is less than $G_B\gamma_B$, the strain $\gamma(t)$ saturates to the value obtained from Eq. (38). By contrast, when $\tau_{app} > G_B\gamma_B$, the strain does not saturate and continues to increase with time. The creep transition is more clearly observed in Figs. 6(c) and 6(d). For $\tau_{app} < G_B\gamma_B$, the shear rate decreases rapidly with time and decays to zero. Otherwise, the shear rate decreases but approaches a nonzero constant, thus implying that the strain continues to increase. In summary, because the Herschel–Bulkley fit yields $\tau_y = G_B \gamma_B$, the yield stress of the Herschel–Bulkley equation provides a reasonable estimate of the threshold for the transition from a viscoelastic solid to a viscoelastic fluid.

# 4. 3D Constitutive Equation

Although the 1D constitutive equation is qualitatively consistent with the experimental data, it is restricted to simple shear. Because practical processes involving yield-stress fluids require predictions under general deformation modes, one must extend the 1D constitutive equation to a 3D formulation. The additive decomposition of strain in Eq. (1) is valid for small deformations, whereas a multiplicative decomposition must be conducted for finite deformations [Gurtin, Fried and Anand, 2010]. Moreover, because a constitutive equation should be invariant under rigid body motions such as translations and rotations, it must satisfy the principle of material-frame invariance.

## 4.1. Theory

According to the Kröner–Lee decomposition [Kröner, 1960; Lee, 1969], the deformation gradient $\mathbf{F}$ is expressed as

$$\mathbf{F} = \mathbf{F}_e \cdot \mathbf{F}_p, \tag{42}$$

where $\mathbf{F}_e$ and $\mathbf{F}_p$ denote the elastic and plastic components, respectively. As the deformation gradient $\mathbf{F}$ is a mapping from the reference configuration to the current configuration (deformed configuration), the decomposition of Eq. (42) implies that an intermediate configuration exists. $\mathbf{F}_p$ is a mapping from the reference configuration to the intermediate configuration, whereas $\mathbf{F}_e$ is a mapping from the intermediate configuration to the current configuration.

Because the velocity gradient $\mathbf{L} = (\nabla \mathbf{v})^T$ is related to $\mathrm{d}\mathbf{F}/\mathrm{d}t = \mathbf{L} \cdot \mathbf{F}$, where $\mathrm{d}/\mathrm{d}t$ is the material time derivative, we have

$$\mathbf{L} = \mathbf{L}_e + \mathbf{L}_p, \tag{43}$$

where

$$\mathbf{L}_e = \frac{\mathrm{d}\mathbf{F}_e}{\mathrm{d}t}\cdot\mathbf{F}_e^{-1};\quad \mathbf{L}_p = \mathbf{F}_e\cdot\hat{\mathbf{L}}_p\cdot\mathbf{F}_e^{-1};\quad \hat{\mathbf{L}}_p = \frac{\mathrm{d}\mathbf{F}_p}{\mathrm{d}t}\cdot\mathbf{F}_p^{-1}. \tag{44}$$

The following relations are widely accepted as the uniqueness of multiplicative decomposition [Gurtin, Fried and Anand, 2010]:

$$\det\mathbf{F} = \det\mathbf{F}_e;\quad \det\mathbf{F}_p = 1;\quad \mathbf{L}_p = \mathbf{L}_p^T \equiv \mathbf{D}_p;\quad \mathrm{tr}\mathbf{L}_p = 0. \tag{45}$$

In analogy to the left Cauchy–Green strain, we express elastic strain as follows:

$$\mathbf{B}_e \equiv \mathbf{F}_e\cdot\mathbf{F}_e^T. \tag{46}$$

In the 3D extension, the total Cauchy stress $\mathbf{T}$ is decomposed as

$$\mathbf{T} = \mathbf{T}_{\mathrm{gel}} + \mathbf{T}_{\mathrm{sol}}, \tag{47}$$

where $\mathbf{T}_{\mathrm{gel}}$ and $\mathbf{T}_{\mathrm{sol}}$ represent the gel and solvent stresses, respectively. Because the gel stress is described by the Zener element, the gel stress $\mathbf{T}_{\mathrm{gel}}$ is a function of $\mathbf{B}_e$. Therefore, the evolution equation of the elastic strain is required, which is expressed as

$$\begin{aligned}\frac{\mathrm{d}\mathbf{B}_e}{\mathrm{d}t} &= \mathbf{L}_e\cdot\mathbf{B}_e + \mathbf{B}_e\cdot\mathbf{L}_e^T \\ &= \mathbf{L}\cdot\mathbf{B}_e + \mathbf{B}_e\cdot\mathbf{L}^T - \mathbf{D}_p\cdot\mathbf{B}_e - \mathbf{B}_e\cdot\mathbf{D}_p,\end{aligned} \tag{48}$$

where $\mathbf{L}_p = \mathbf{D}_p$ is assumed. Next, the plastic deformation rate $\mathbf{D}_p$ is to be modeled constitutively. According to Leonov (1976), the plastic deformation rate is expressed as follows:

$$\mathbf{D}_p\cdot\mathbf{B}_e + \mathbf{B}_e\cdot\mathbf{D}_p = \frac{1}{\eta}\left(\mathbf{T}'_{\mathrm{gel}} - \mathbf{T}'_B\right), \tag{49}$$

where the prime " $'$ " denotes the deviatoric component of a second-order tensor,

$$\mathbf{A}' \equiv \mathbf{A} - \frac{\mathrm{tr}\mathbf{A}}{3}\mathbf{I}, \tag{50}$$

and $\mathbf{T}_B$ is the back stress, which is exerted on the spring of the Kelvin–Voigt element. Therefore, $\mathbf{T}_{\mathrm{gel}} - \mathbf{T}_B$ is the stress exerted on the dashpot of the Kelvin–Voigt element. Meanwhile, Tervoort, Klompen, and Govaert (1996) suggested another model for $\mathbf{D}_p$, i.e.,

$$\mathbf{D}_p = \frac{1}{\eta}\left(\mathbf{T}'_{\text{gel}} - \mathbf{T}'_B\right). \tag{51}$$

Equations (49) and (51) are known as flow rules in plasticity terminology. To select between the two flow rules, one must consider the back-stress evolution equation. The original Armstrong–Frederick equation does not satisfy the principle of material-frame invariance because it is expressed as by

$$\frac{\mathrm{d}\hat{\mathbf{T}}_B}{\mathrm{d}t} = G_B\hat{\mathbf{D}}_p - \frac{\dot{e}_p}{\gamma_B}\hat{\mathbf{T}}_B, \tag{52}$$

where

$$\dot{e}_p = \sqrt{2\mathbf{D}_p : \mathbf{D}_p}. \tag{53}$$

Because the back stress is exerted on the spring of the Kelvin-Voigt element, the back stress $\hat{\mathbf{T}}_B$ in Eq. (52) must be a linear mapping from a vector in the intermediate configuration to a vector in the same configuration. By contrast, the back stress $\mathbf{T}_B$ in Eqs. (49) and (51) is a linear mapping from a vector of the current configuration to another vector of the current configuration; therefore, it is different from the back stress $\hat{\mathbf{T}}_B$ in Eq. (52). The same reasoning can be applied to the plastic deformation rate $\hat{\mathbf{D}}_p$ in Eq. (52) and to the plastic deformation rates $\mathbf{D}_p$ in Eqs. (49) and (51). Therefore, one may consider the following relation:

$$\mathbf{T}_B = \mathbf{F}_e \cdot \hat{\mathbf{T}}_B \cdot \mathbf{F}_e^T. \tag{54}$$

This is analogous to the relation between the Cauchy and Piola stresses [Lai, Rubin and Krempl, 2010]. Taking the material time derivative of Eq. (54) and applying Eq. (52), we obtain

$$\frac{\mathrm{d}\mathbf{T}_B}{\mathrm{d}t} = \mathbf{L}\cdot\mathbf{T}_B + \mathbf{T}_B\cdot\mathbf{L}^T - \mathbf{D}_p\cdot\mathbf{T}_B - \mathbf{T}_B\cdot\mathbf{D}_p + G_B\mathbf{F}_e\cdot\hat{\mathbf{D}}_p\cdot\mathbf{F}_e^T - \frac{\dot{e}_p}{\gamma_B}\mathbf{T}_B, \tag{55}$$

where Eq. (43), Eq. (44), and the assumption of $\mathbf{L}_p = \mathbf{D}_p$ are used. If we consider $\hat{\mathbf{D}}_p$ as the symmetric component of $\hat{\mathbf{L}}_p$, then of $\hat{\mathbf{L}}_p$, then Eq. (55) is expressed as

$$\frac{\mathrm{d}\mathbf{T}_B}{\mathrm{d}t} = \mathbf{L}\cdot\mathbf{T}_B + \mathbf{T}_B\cdot\mathbf{L}^T - \mathbf{D}_p\cdot\left(\mathbf{T}_B - \tfrac{1}{2}G_B\mathbf{B}_e\right) - \left(\mathbf{T}_B - \tfrac{1}{2}G_B\mathbf{B}_e\right)\cdot\mathbf{D}_p - \frac{\dot{e}_p}{\gamma_B}\mathbf{T}_B, \tag{56}$$

because

$$\begin{aligned}\mathbf{F}_e\cdot\hat{\mathbf{D}}_p\cdot\mathbf{F}_e^T &= \frac{1}{2}\mathbf{F}_e\cdot\left(\hat{\mathbf{L}}_p + \hat{\mathbf{L}}_p^T\right)\cdot\mathbf{F}_e^T \\ &= \frac{1}{2}\mathbf{F}_e\cdot\left(\mathbf{F}_e^{-1}\cdot\mathbf{L}_p\cdot\mathbf{F}_e + \mathbf{F}_e^T\cdot\mathbf{L}_p^T\cdot\mathbf{F}_e^{-T}\right)\cdot\mathbf{F}_e^T \\ &= \frac{1}{2}\left(\mathbf{D}_p\cdot\mathbf{B}_e + \mathbf{B}_e\cdot\mathbf{D}_p\right).\end{aligned} \tag{57}$$

Instead of $\hat{\mathbf{D}}_p = \frac{1}{2}\left(\hat{\mathbf{L}}_p + \hat{\mathbf{L}}_p^T\right)$, one may consider the following expression for $\hat{\mathbf{D}}_p$:

$$\hat{\mathbf{D}}_p \equiv \mathbf{F}_e^{-1}\cdot\mathbf{D}_p\cdot\mathbf{F}_e^{-T} \Leftrightarrow \mathbf{D}_p = \mathbf{F}_e\cdot\hat{\mathbf{D}}_p\cdot\mathbf{F}_e^T. \tag{58}$$

Thus, the evolution equation of $\mathbf{T}_B$ is expressed as

$$\frac{\mathrm{d}\mathbf{T}_B}{\mathrm{d}t} = \mathbf{L}\cdot\mathbf{T}_B + \mathbf{T}_B\cdot\mathbf{L}^T - \mathbf{D}_p\cdot\mathbf{T}_B - \mathbf{T}_B\cdot\mathbf{D}_p + G_B\mathbf{D}_p - \frac{\dot{e}_p}{\gamma_B}\mathbf{T}_B. \tag{59}$$

Here, we use Eq. (58) as using Eq. $(44)_2$ is not consistent with the assumption $\mathbf{L}_p = \mathbf{D}_p$ because $\hat{\mathbf{D}}_p = \mathbf{F}_e^{-1}\cdot\mathbf{D}_p\cdot\mathbf{F}_e$ does not guarantee the symmetry of $\hat{\mathbf{D}}_p$ even though $\mathbf{D}_p$ is symmetric. Hence, we model the evolution equation of back stress as shown in Eq. (59).

Next, we must select the flow rule between Eqs. (49) and (51). If we select Eq. (49), which is an implicit equation of $\mathbf{D}_p$, then, we must ascertain an analytical solution to the following equation for tensor algebra:

$$\mathbf{S}\cdot\mathbf{X} + \mathbf{X}\cdot\mathbf{S} = \mathbf{Z}, \tag{60}$$

where $\mathbf{S}$ is known as an invertible symmetric tensor. $\mathbf{Z}$ is known, whereas $\mathbf{X}$ is to be obtained: $\mathbf{S} = \mathbf{B}_e$, $\mathbf{X} = \mathbf{D}_p$, and $\mathbf{Z} = \left(\mathbf{T}'_{\mathrm{gel}} - \mathbf{T}'_B\right)/\eta$ for Eq.(49). This problem has been extensively investigated [Scheidler, 1994; Ting, 1996; Rosati, 2000]. The solution to Eq. (61) is expressed as

$$\begin{aligned}2\left(I_{\mathrm{S}}II_{\mathrm{S}} - III_{\mathrm{S}}\right)\mathbf{X} = {} & \left(I_{\mathrm{S}}^2 + II_{\mathrm{S}}\right)\mathbf{Z} - I_{\mathrm{S}}\left(\mathbf{S}\cdot\mathbf{Z} + \mathbf{Z}\cdot\mathbf{S}\right) - III_{\mathrm{S}}\left(\mathbf{S}^{-1}\cdot\mathbf{Z} + \mathbf{Z}\cdot\mathbf{S}^{-1}\right) \\ & + \mathbf{S}\cdot\mathbf{Z}\cdot\mathbf{S} + I_{\mathrm{S}}III_{\mathrm{S}}\mathbf{S}^{-1}\cdot\mathbf{Z}\cdot\mathbf{S}^{-1},\end{aligned} \tag{61}$$

where $I_S$, $II_S$, and $III_S$ are the principal invariants of $\mathbf{S}$. Therefore, applying Eq. (49) to Eq. (59) requires a cumbersome algebraic manipulation. However, using Eq. (51) offers more convenience than using Eq. (49).

The next step in constitutive modeling is to model the viscosity $\eta$. The Eyring and Carreau–Yasuda viscosity models were adopted in a manner consistent with the 1D model, such that

$$\begin{aligned} \eta_{\mathrm{Eyr}} &= \eta_o \frac{\xi}{\sinh \xi} \quad \text{Eyring viscosity;} \\ \eta_{\mathrm{CY}} &= \frac{\eta_o}{\left(1+\xi^{\mu}\right)^{\nu}} \quad \text{Carreau-Yasuda viscosity,} \end{aligned} \tag{62}$$

where

$$\xi = \frac{\langle \tau \rangle}{\tau_o}, \quad \langle \tau \rangle \equiv \sqrt{\frac{1}{2}\left(\mathbf{T}_{\mathrm{gel}} - \mathbf{T}_B\right)' : \left(\mathbf{T}_{\mathrm{gel}} - \mathbf{T}_B\right)'}. \tag{63}$$

The Kröner-Lee decomposition is based on the concepts of the reference space (reference configuration), structural space (intermediate configuration) and observed space (current configuration) [Gurtin, Fried and Anand, 2010]. A vector in the reference space is designated as a material vector, a vector in the structural space is termed a structural vector, and a vector in the observed space is referred to as a spatial vector. The plastic deformation gradient $\mathbf{F}_p$ is a linear mapping from a material vector to a structural vector and the elastic deformation tensor $\mathbf{F}_e$ is a linear mapping from a structural vector to a spatial vector. Meanwhile, the velocity gradient $\mathbf{L}$ is a linear mapping between two spatial vectors. Since the left and the right-hand sides of Eq. (43) should belong to the same tensor category, both $\mathbf{L}_e$ and $\mathbf{L}_p$ are linear mappings between two spatial vectors, which are known as spatial tensors. Since $\mathbf{F}_e$ is a mapping from a structural vector to a spatial vector, its inverse $\mathbf{F}_e^{-1}$ is a mapping from a spatial vector to a structural vector and $\mathrm{d}\mathbf{F}_e/\mathrm{d}t$ belongs to the category of $\mathbf{F}_e$. Therefore, $\mathbf{L}_e$ is a linear mapping between two vectors in the observed space. The same reasoning shows that $\mathbf{L}_p$ belongs to the category of $\mathbf{L}$. Clearly, $\hat{\mathbf{L}}_p$ is a linear mapping between two structural vectors (structural tensor).

Considering the uniqueness of the Kröner–Lee decomposition, we can reasonably assume that the structural tensors are invariant with respect to the observers who conduct rigid body motion, similarly to the material tensors, which are a linear mapping from the material space to itself. Moreover, the spatial tensors must obey

$$\mathbf{A}^* = \mathbf{Q} \cdot \mathbf{A} \cdot \mathbf{Q}^T, \tag{64}$$

where $\mathbf{A}$ is an arbitrary spatial tensor observed by a fixed observer, $\mathbf{A}^*$ is the tensor $\mathbf{A}$ observed by a moving observer and $\mathbf{Q}$ is an orthogonal tensor representing the rotational motion of the moving observer. Here, $\mathbf{T}_{\text{gel}}$, $\mathbf{T}_B$, and $\mathbf{B}_e$ are spatial tensors. Clearly, the evolution equations of the spatial tensors obey the principle of material-frame invariance.

The final block of the constitutive equations is the relation between the elastic strain $\mathbf{B}_e$ and gel stress $\mathbf{T}_{\text{gel}}$. A hyperelasticity model, such as the BST model [Blatz, Sharda and Tschoegl, 1974] can be adopted. Nonetheless, the subsequent relation is regarded as one of the simplest, i.e.,

$$\mathbf{T}_{\text{gel}} = -p\mathbf{I} + G\mathbf{B}_e', \tag{65}$$

where $p$ denotes the pressure. Equation (65) was used by Tervoort, Klompen and Govaert (1996) and is a simple modification of the infinitesimal linear elasticity. In this study, we selected the following solvent stress:

$$\mathbf{T}_{\text{sol}} = 2\eta_s \mathbf{D}'. \tag{66}$$

In principle, the pressure $p$ is determined by the momentum balance and boundary conditions. However, in simple shear, $p$ contributes only to isotropic terms and is therefore irrelevant to the shear stress component $T_{12}$ or to the normal stress differences. Moreover, both the viscosity and plastic deformation rate $\mathbf{D}_p$ are defined from the deviatoric component of stress; thus, $p$ does not affect their evaluation. Accordingly, when necessary, $p$ may be established by selecting conditions such as $T_{33} = 0$. However, under general flow conditions, $p$ must be selected by coupling the momentum balance with the appropriate boundary conditions. Because this study focused on a simple shear flow, the pressure need not be considered.

Consequently, our constitutive equation results in the relations:

$$
\begin{aligned}
&\frac{\mathrm{d}\mathbf{B}_e}{\mathrm{d}t} = \mathbf{L}\cdot\mathbf{B}_e + \mathbf{B}_e\cdot\mathbf{L}^T - \mathbf{D}_p\cdot\mathbf{B}_e - \mathbf{B}_e\cdot\mathbf{D}_p;\\
&\mathbf{D}_p = \frac{1}{\eta}\left(\mathbf{T}'_{\mathrm{gel}} - \mathbf{T}'_B\right);\\
&\eta = \begin{cases} \eta_o \dfrac{\xi}{\sinh\xi} & \text{Eyring viscosity}\\ \dfrac{\eta_o}{\left(1+\xi^{\mu}\right)^{\nu}} & \text{Carreau-Yasuda viscosity;}\end{cases}\\
&\xi = \frac{\langle\tau\rangle}{\tau_o},\quad \langle\tau\rangle \equiv \sqrt{\frac{1}{2}\left(\mathbf{T}_{\mathrm{gel}} - \mathbf{T}_B\right)' : \left(\mathbf{T}_{\mathrm{gel}} - \mathbf{T}_B\right)'};\\
&\frac{\mathrm{d}\mathbf{T}_B}{\mathrm{d}t} = \mathbf{L}\cdot\mathbf{T}_B + \mathbf{T}_B\cdot\mathbf{L}^T - \mathbf{D}_p\cdot\mathbf{T}_B - \mathbf{T}_B\cdot\mathbf{D}_p + G_B\mathbf{D}_p - \frac{\dot{e}_p}{\gamma_B}\mathbf{T}_B;\\
&\dot{e}_p = \sqrt{2\mathbf{D}_p:\mathbf{D}_p};\\
&\mathbf{T} = \mathbf{T}_{\mathrm{gel}} + \mathbf{T}_{\mathrm{sol}};\quad \mathbf{T}_{\mathrm{gel}} = -p\mathbf{I} + G\mathbf{B}'_e;\quad \mathbf{T}_{\mathrm{sol}} = 2\eta_s\mathbf{D}'.
\end{aligned}
\tag{67}
$$

The present model has several features in common with the plasticity-inspired KH approach [Dimitriou and McKinley, 2019], including a flow rule, back-stress evolution based on the Armstrong–Frederick theory, and a 3D extension using Kröner–Lee decomposition, despite the differences in the details of the derivation process. However, in contrast to the KH framework, the present constitutive equation is derived from an explicit viscoelastic-solid interpretation of simple yield-stress fluids, motivated by the recent experimental results to show the non-zero fully relaxed stress in Carbopol dispersions [Vinutha *et al.*, 2024]. In addition, the model incorporates a Zener-type gel phase response, in conjunction with a distinct parallel solvent dissipation pathway.

## 4.2. Model Calculation of 3D Constitutive Equation

To facilitate understanding, Eq. (67) is non-dimensionalized as follows:

$$
\begin{aligned}
&\frac{\mathrm{d}\mathbf{B}_e}{\mathrm{d}\tilde{t}} = \tilde{\mathbf{L}}\cdot\mathbf{B}_e + \mathbf{B}_e\cdot\tilde{\mathbf{L}}^T - \tilde{\mathbf{D}}_p\cdot\mathbf{B}_e - \mathbf{B}_e\cdot\tilde{\mathbf{D}}_p;\\
&\tilde{\mathbf{D}}_p = \frac{1}{\tilde{\eta}}\left(\tilde{\mathbf{T}}'_{\mathrm{gel}} - \tilde{\mathbf{T}}'_B\right);\\
&\frac{\mathrm{d}\tilde{\mathbf{T}}_B}{\mathrm{d}\tilde{t}} = \tilde{\mathbf{L}}\cdot\tilde{\mathbf{T}}_B + \tilde{\mathbf{T}}_B\cdot\tilde{\mathbf{L}}^T - \tilde{\mathbf{D}}_p\cdot\tilde{\mathbf{T}}_B - \tilde{\mathbf{T}}_B\cdot\tilde{\mathbf{D}}_p + \tilde{G}_B\tilde{\mathbf{D}}_p - \frac{\tilde{\dot{e}}_p}{\gamma_B}\tilde{\mathbf{T}}_B;\\
&\tilde{\dot{e}}_p = \sqrt{2\tilde{\mathbf{D}}_p:\tilde{\mathbf{D}}_p};\\
&\tilde{\mathbf{T}} = -\tilde{p}\mathbf{I} + \tilde{G}\mathbf{B}'_e + 2\tilde{\eta}_s\tilde{\mathbf{D}}',
\end{aligned}
\tag{68}
$$

where

$$\lambda \equiv \frac{\eta_o}{\tau_o}; \quad \tilde{t} \equiv \frac{t}{\lambda}; \quad \tilde{\mathbf{L}} \equiv \lambda \mathbf{L}; \quad \tilde{\mathbf{D}}_p \equiv \lambda \mathbf{D}_p; \quad \tilde{\eta} \equiv \frac{\eta}{\eta_o};$$
$$\tilde{\mathbf{T}}_{gel} \equiv \frac{1}{\tau_o} \mathbf{T}_{gel}; \quad \tilde{\mathbf{T}}_B \equiv \frac{1}{\tau_o} \mathbf{T}_B; \quad \tilde{G}_B \equiv \frac{G_B}{\tau_o}; \quad \tilde{\dot{e}}_p \equiv \lambda \dot{e}_p; \tag{69}$$
$$\tilde{\mathbf{T}} \equiv \frac{1}{\tau_o} \mathbf{T}; \quad \tilde{p} \equiv \frac{p}{\tau_o}; \quad \tilde{G} \equiv \frac{G}{\tau_o}; \quad \tilde{\eta}_s \equiv \frac{\eta_s}{\eta_o}; \quad \tilde{\mathbf{D}} \equiv \lambda \mathbf{D}.$$

The Eyring and Carreau–Yasuda viscosity models (Eq. (67)3) are adopted for the viscosity.

Using the 3D constitutive equation, we computed the start-up shear, stress relaxation and creep as in the 1D case. The parameters were set as follows: $\gamma_B = 0.1$, $G = 20\,\text{Pa}$, $G_B = 100\,\text{Pa}$, $\tau_o = 30\,\text{Pa}$, $\eta_o = 70\,\text{Pa}\cdot\text{s}$, $\eta_s = 1\,\text{Pa}\cdot\text{s}$, $\mu = 2.5$, $\nu = 1.5$ and $\lambda = \eta_o/\tau_o$. To examine the steady state stress, we calculated the start-up shear until a total strain of $\gamma = 20$. For the start-up shear, the initial conditions of $\mathbf{B}_e$ and $\mathbf{T}_B$ were $\mathbf{I}$ and $\mathbf{O}$, respectively. Next, the subsequent stress relaxation was calculated for the same time as in the calculations of the start-up shear such that the stress had sufficient time to approach the fully relaxed stress.

For the creep simulations, we imposed a step shear stress $T_{12}(t) = T_{12}^{\text{app}}$ for $t > 0$. Since it is not straightforward to prescribe initial conditions that are consistent with a step stress in the 3D formulation, we approximated the onset of creep by a short start-up segment. Specifically, starting from the undeformed state $\mathbf{B}_e(0) = \mathbf{I}$ and $\mathbf{T}_B(0) = \mathbf{O}$, we performed short start-up shear calculations over $0 \le t \le \lambda/100$ using trial shear rates $\dot{\gamma}_o$. The trial shear rate was adjusted until the instantaneous shear stress satisfies $T_{12}(\lambda/100) = T_{12}^{\text{app}}$. The values $\mathbf{B}_e(\lambda/100)$ and $\mathbf{T}_B(\lambda/100)$ obtained from this short start-up segment were then used as the initial conditions for the subsequent stress-controlled creep simulation. The imposed shear stress $T_{12}^{\text{app}}$ was maintained by adjusting the shear rate. The corresponding shear rate was determined at each time step as follows:

$$\frac{d\gamma}{dt} = \frac{T_{12}^{\text{app}} - GB_{12}^{(e)}}{\eta_s}. \tag{70}$$

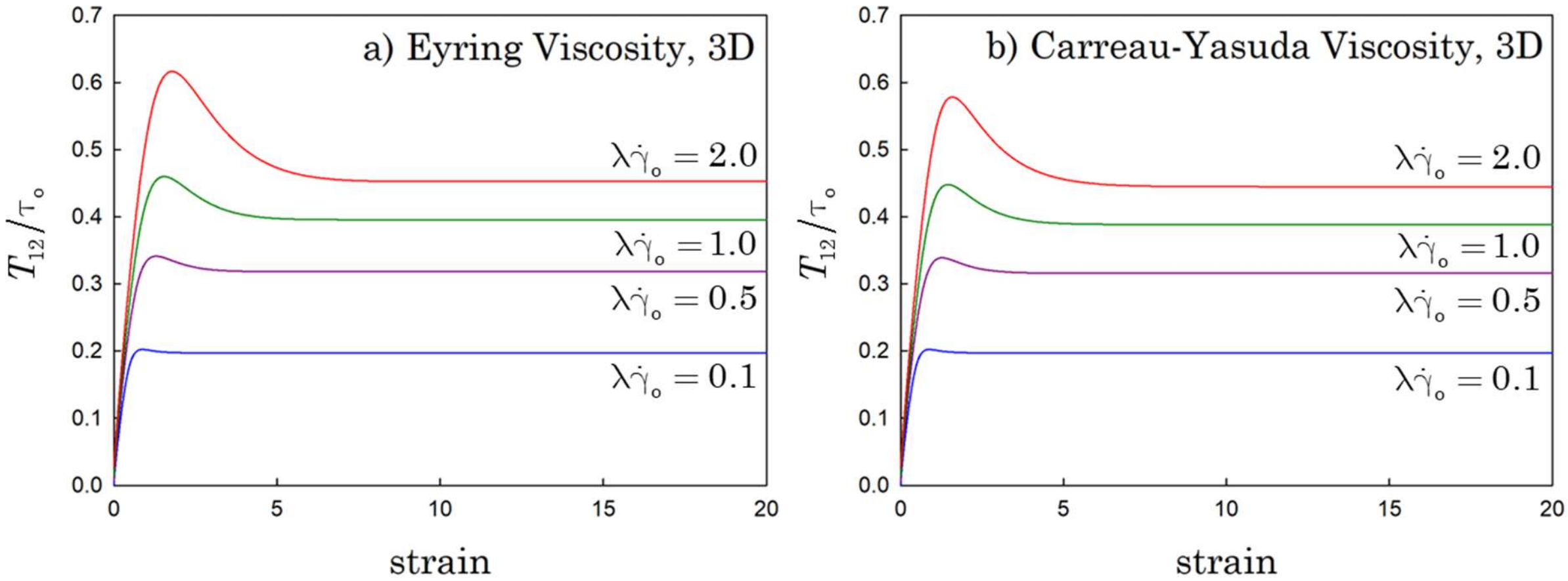

**Fig. 7.** Shear stresses of start-up shear calculated via numerical integration of Eq. (68) using (a) Eyring viscosity and (b) Carreau–Yasuda viscosity. The parameters are set as follows: $\gamma_B = 0.1$, $G = 20\,\text{Pa}$, $G_B = 100\,\text{Pa}$, $\tau_o = 30\,\text{Pa}$, $\eta_o = 70\,\text{Pa}\cdot\text{s}$, $\eta_s = 1\,\text{Pa}\cdot\text{s}$, $\mu = 2.5$, $\nu = 1.5$ and $\lambda = \eta_o/\tau_o$.

Figure 7 shows the start-up shear responses obtained by numerically integrating the 3D constitutive equation (Eq. (68)) at a constant shear rate. As in the 1D model, the shear stress increased with time (or strain) and approached a steady-state value. As the imposed shear rate $\dot{\gamma}_o$ increased, the steady-state stress $T_{12}(t \to \infty)$ increased for the Eyring and Carreau–Yasuda viscosities. Moreover, the strain required to reach the steady state increased with the shear rate. In contrast to the 1D constitutive equation, the 3D model exhibited a stress overshoot in the start-up shear. The overshoot became more pronounced as the shear rate increased, which agreed qualitatively with the experimental observations reported by Pagani *et al.* (2024).

To elucidate the origin of the overshoot, we analyze the rate of the gel stress rather than the stress itself. Since the imposed strain rate is constant in the start-up flow, the solvent stress remains constant; hence, the observed overshoot originates primarily from the transient evolution of the gel stress. In the 1D model, $\dot{\gamma}_e = \dot{\gamma}_o - \dot{\gamma}_p$ yields $d\tau_{\text{gel}}/dt = G\left(\dot{\gamma}_o - \dot{\gamma}_p\right)$. Therefore, a stress overshoot requires a finite strain interval in which the relaxation term exceeds the elastic loading term, i.e., $G\dot{\gamma}_p > G\dot{\gamma}_o$. However, the analytical argument based on Eq. (21) shows that the corresponding solution is unique, which suppresses the formation of such a transient overtaking regime and results in a monotonic convergence to a steady value.

To make this competition explicit, we define an auxiliary function,

$$
\begin{aligned}
& dT_{12}^{(\mathrm{gel})} \big/ dt = E_{3D} - R_{3D}; \\
& E_{3D} \equiv G B_{22}^{(e)} \dot{\gamma}_{o}; \\
& R_{3D} \equiv G\left\{ \left[ D_{11}^{(p)} + D_{22}^{(p)} \right] B_{12}^{(e)} - D_{12}^{(p)} \left[ B_{11}^{(e)} + B_{22}^{(e)} \right] \right\},
\end{aligned} \tag{71}
$$

where $E_{3D}$ and $R_{3D}$ denote the elastic loading and relaxation terms for the 3D model, respectively. In the 1D model, gel stress rate $d\tau_{\mathrm{gel}}/dt$ remains positive over the finite strain range considered and approaches zero only asymptotically with increasing strain. This indicates that the relaxation term never exceeds the elastic loading term at finite strain, and therefore no stress overshoot occurs. In contrast, $dT_{12}^{(\mathrm{gel})}/dt$ value crosses zero at a finite strain in the 3D model. The first zero of $dT_{12}^{(\mathrm{gel})}/dt$, or the first crossing of the elastic and relaxation terms, coincides with the stress peak. Above the crossing point, $dT_{12}^{(\mathrm{gel})}/dt < 0$, which means that the relaxation term temporarily exceeds the elastic loading term, causing the stress to decrease. The subsequent return of $dT_{12}^{(\mathrm{gel})}/dt$ toward zero at larger strain corresponds to the approach to the steady state. Plots of the elastic loading term, relaxation term, and gel stress rate for the 1D and 3D models are provided in Supplementary material.

The emergence of the stress overshoot in the 3D start-up shear can be attributed to the tensorial structure introduced in the 3D model. In the 3D formulation, the plastic deformation rate $\mathbf{D}_p$ is governed by Eq. (67)2, whereas the viscosity is dependent upon $\xi$, which is defined by the second invariant of the deviatoric stress difference $\mathbf{T}'_{\mathrm{gel}} - \mathbf{T}'_B$ (Eqs. (67)4 and (67)5). It has been established that normal stress components develop even under simple shear in the 3D formations. These components contribute to $\xi$, thereby accelerating the rate of shear thinning of $\eta(\xi)$ and, consequently, transiently amplifying $|\mathbf{D}_p|$. Therefore, the relaxation term can temporarily exceed the elastic loading term, leading to a stress overshoot in the total shear stress ($T_{12}$) before the stress reaches a steady state. These observations demonstrate that the overshoot is primarily governed by the tensorial structure of the 3D constitutive equation, rather than by the specific choice of the viscosity law itself. It has been previously postulated that stress overshoot can occur as isotropic hardening [Dimitriou and McKinley, 2014] or spatial microstructural inhomogeneity [Benzi *et al.*, 2021] develops. However, the current findings demonstrate that the phenomenon can be predicted from viscoelastic solid-based constitutive equations without such additional assumptions. This finding indicates that the isotropic hardening [Dimitriou and McKinley, 2014] or the spatial microstructural inhomogeneity [Benzi *et al.*, 2021] may not be a prerequisite for the occurrence of the stress overshoot in the start-up shear.

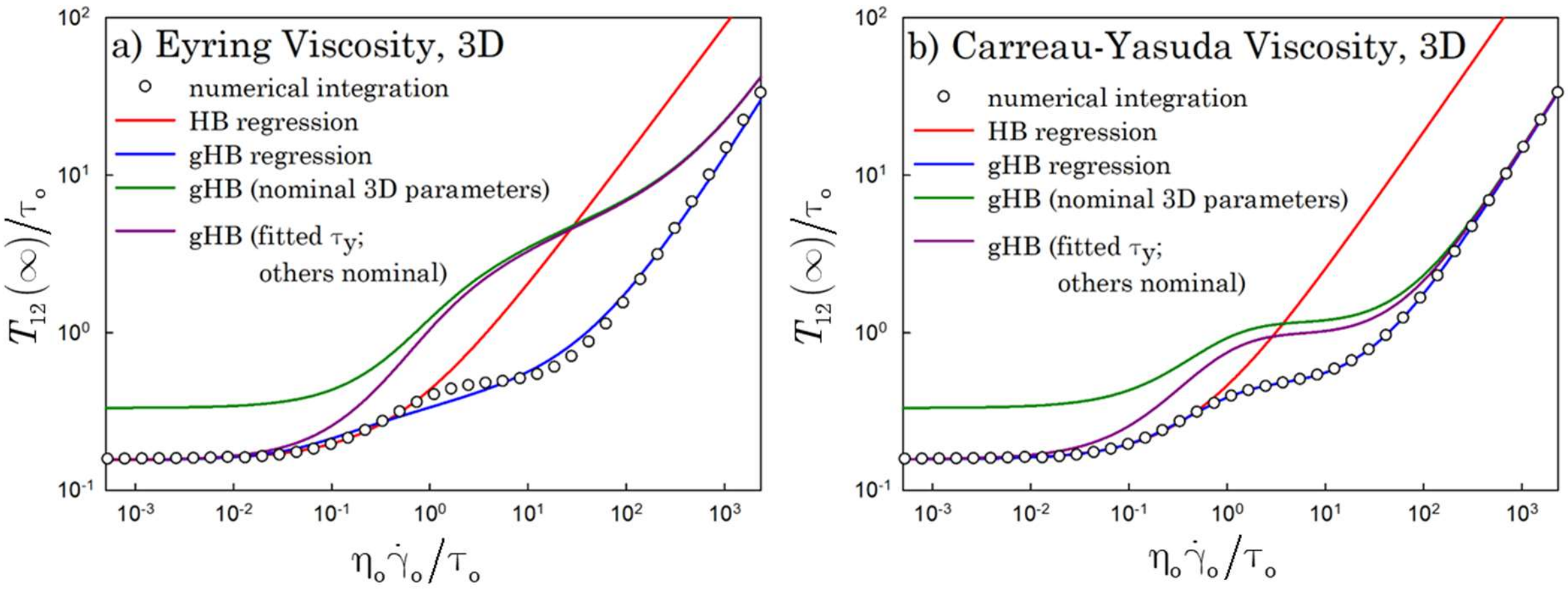

**Fig. 8.** Steady state stresses calculated using (a) Eyring viscosity and (b) Carreau-Yasuda viscosity plotted as a function of normalized shear rate $\eta_o\dot{\gamma}_o/\tau_o$. Open circles represent steady state stresses obtained from numerical integration of Eq. (68); red lines represent regression results of steady-state stress obtained using Herschel–Bulkley equation (HB regression); blue lines represent regression results using generalized Herschel-Bulkley equation (gHB regression); green lines represent the corresponding generalized Herschel-Bulkley curve with the nominal parameters used in the 3D simulations (gHB (nominal 3D parameters)); and the purple lines represent the generalized Herschel-Bulkley curve evaluated with the fitted $\tau_y$ while keeping the remaining parameters at their nominal 3D values (gHB (fitted $\tau_y$; others nominal)). For the Carreau-Yasuda case, $\alpha$ and $\beta$ appearing in the gHB equations are taken from the gHB regression. The parameters for the 3D simulations are set as follows: $\gamma_B = 0.1$, $G = 20\,\mathrm{Pa}$, $G_B = 100\,\mathrm{Pa}$, $\tau_o = 30\,\mathrm{Pa}$, $\eta_o = 70\,\mathrm{Pa \cdot s}$, $\eta_s = 1\,\mathrm{Pa \cdot s}$, $\mu = 2.5$, $\nu = 1.5$ and $\lambda = \eta_o/\tau_o$.

**Table 1.** Parameters obtained from generalized Herschel–Bulkley regressions to the steady-state stresses computed by the 3D model in Fig . 8.

| Parameters | Eyring viscosity | Carreau-Yasuda viscosity |
|---|---|---|
| $\tau_y$ (Pa) | 4.667 | 4.732 |
| $\eta_o$ $(\mathrm{Pa \cdot s})$ | 48.390 | 27.661 |
| $\tau_o$ (Pa) | 1.528 | 10.804 |
| $\eta_s$ $(\mathrm{Pa \cdot s})$ | 0.8780 | 0.9920 |
| $\alpha$ $(-)$ | - | 1.428 |

| $\beta$ $(-)$ | - | 0.789 |
|---|---|---|

Figure 8 shows the steady state stress obtained by numerically integrating Eq. (68) under start-up shear, together with fits to the Herschel-Bulkley equation (also refer to Appendix B). The fitted parameter values of generalized Herschel-Bulkley equations are listed in Table 1. Notably, the fitted yield stresses are both close to 4.7 Pa for gHB regression of Eyring and Carreau-Yasuda viscosities, and the fitted $\eta_s$ remains close to the prescribed solvent viscosity. By contrast, the remaining parameters do not coincide with the nominal values used in the 3D simulations. Therefore, a one-to-one correspondence between the two sets of parameters is not implied.

In both Figs. 8(a) and 8(b), the steady state stress reflects the Herschel-Bulkley relation at low shear rates, whereas deviations become apparent at higher shear rates. In the limit $\dot{\gamma}_o \to 0$, the steady state stress approaches a constant value, which we denote by the yield stress $\tau_y$ obtained from the HB regression. In the 1D constitutive equation, the yield stress satisfies $\tau_y = G_B \gamma_B$, however, in the 3D constitutive equation, $G_B \gamma_B$ is approximately twice the Herschel–Bulkley yield stress for the parameter set considered here, i.e., $G_B \gamma_B \approx 2\tau_y$. This motivates examining which stress scale, $\tau_y$ or $G_B \gamma_B$, better correlates with the threshold stress for the creep transition in the 3D simulations. This will be discussed in Fig. 10.

Figure 8(a) shows that, for the Eyring viscosity, the gHB regression using Eyring viscosity captures the overall trend of the numerical steady state stress and reproduces the low and high shear rate asymptotic behaviors to some extent. However, it does not fully reproduce the change in curvature in the intermediate shear rate regime. The gHB curves evaluated with the nominal 3D parameters, with or without replacing $\tau_y$ by its fitted value, deviate even more substantially from the numerical data in this region. These results indicate that the discrepancy cannot be attributed solely to the effective yield-stress scale and that, when the 3D Eyring-viscosity model is reduced to an Eyring-form gHB representation, the remaining parameters such as $\eta_o$ and $\tau_o$ must also be re-identified as effective parameters. More importantly, they suggest that the Eyring-form gHB itself has limited flexibility in reproducing the curvature change of the 3D steady flow curve.

By contrast, in Fig 8(b), the gHB regression using the Carreau-Yasuda viscosity reproduces the numerical steady state stress almost quantitatively. This behavior is likely related to the greater flexibility of the Carreau–Yasuda viscosity. When the nominal parameters are used, the resulting gHB curve captures the high hear rate regime reasonably well, while replacing $\tau_y$ by its fitted value additionally improves the agreement in the low shear rate regime. Nevertheless, a noticeable discrepancy remains in the intermediate shear-rate regime, showing that correcting only the yield-stress scale is still insufficient to recover the full steady flow curve. Since $\alpha$ and $\beta$ are already taken from the regression in the Carreau–Yasuda case, the remaining mismatch implies that the nominal values of $\eta_o$ and $\tau_o$ are not sufficient to reproduce the crossover region without further effective re-identification. Thus, the greater flexibility of the Carreau–Yasuda form is advantageous in capturing the low- and high-shear asymptotes, but an effective parameter identification is still required to describe the intermediate-shear-rate curvature accurately.

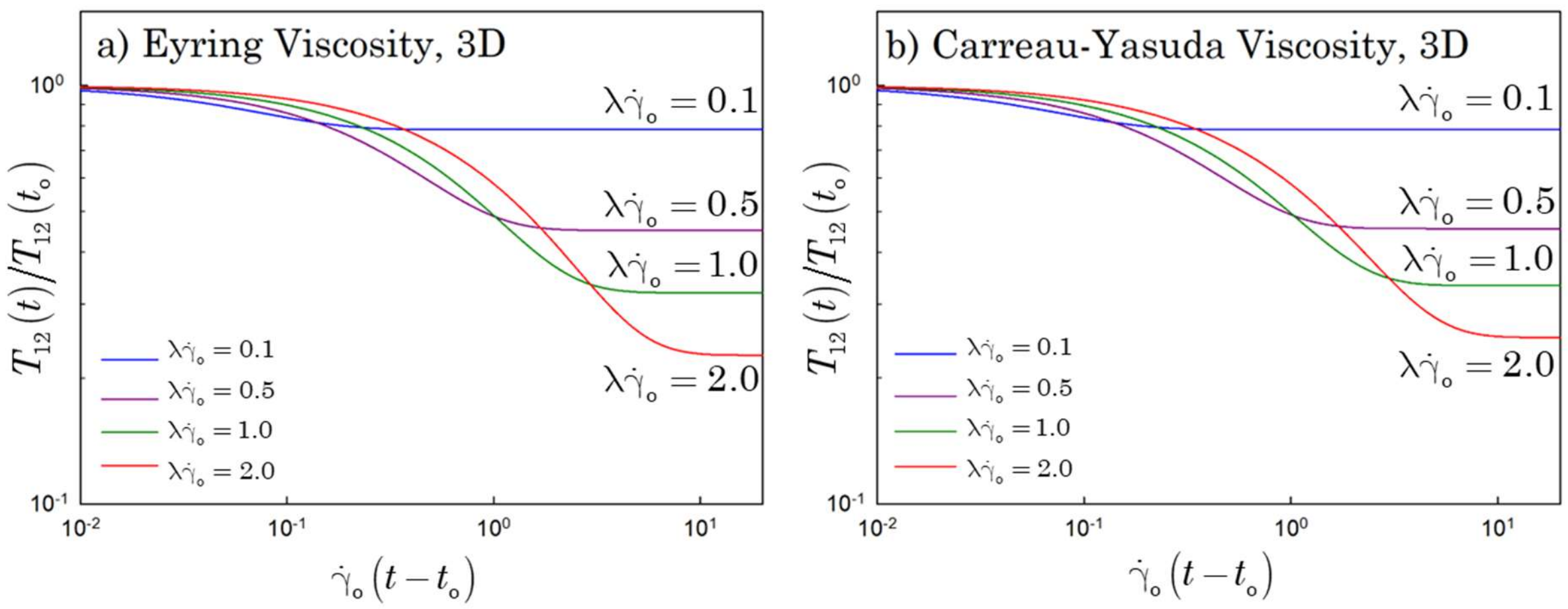

**Fig. 9.** Shear stresses of stress relaxation calculated using (a) Eyring viscosity and (b) Carreau-Yasuda viscosity. The parameters are set as follows: $t_o = 20/\dot{\gamma}_o$, $t_{max} = 40/\dot{\gamma}_o$, $\gamma_B = 0.1$, $G = 20\,\text{Pa}$, $G_B = 100\,\text{Pa}$, $\tau_o = 30\,\text{Pa}$, $\eta_o = 70\,\text{Pa}\cdot\text{s}$, $\eta_s = 1\,\text{Pa}\cdot\text{s}$, $\mu = 2.5$, $\nu = 1.5$ and $\lambda = \eta_o/\tau_o$.

Figure 9 shows the stress relaxation results obtained by numerically integrating Eq. (68) (also refer to Appendix B). At long times, the shear stress approaches a non-zero fully relaxed stress. Normalizing the shear stress by its value at the onset of relaxation $T_{12}(t_o)$, the ratio $T_{12}(t)/T_{12}(t_o)$ exhibits a smaller long-time plateau as the pre-shearing rate $\dot{\gamma}_o$ increases. In contrast to the results calculated from the 1D constitutive equation, the fully relaxed stress calculated from the 3D model is not fixed at $T_{12}(\infty) = G_B\gamma_B$ but decreases with increasing pre-shearing rate. That is, in the 3D constitutive equation, the fully relaxed stress becomes a function of the pre-shearing rate.

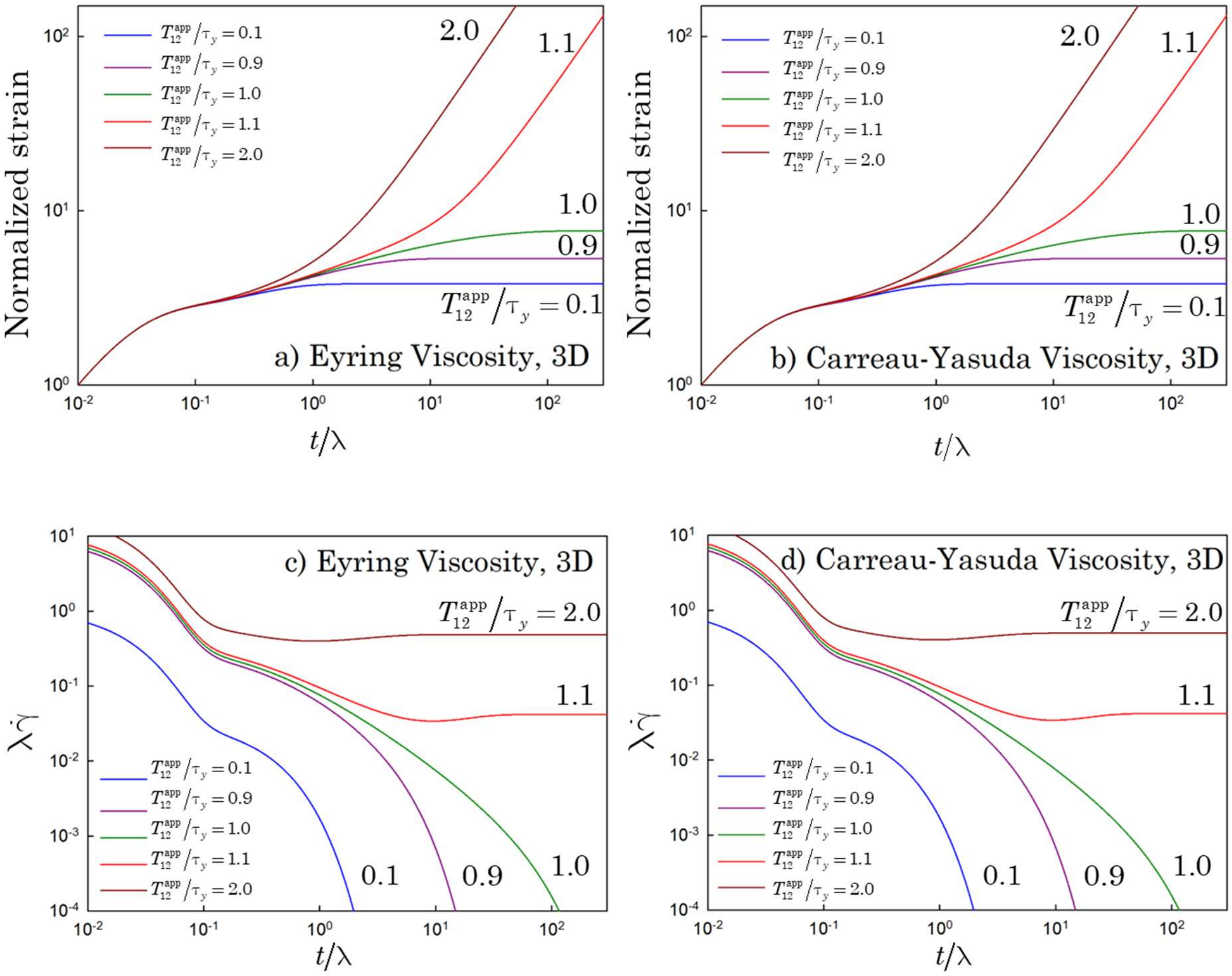

**Fig. 10.** Numerical integration results of Eq. (68) for creep test. Strains calculated using (a) Eyring viscosity and (b) Carreau-Yasuda viscosity and shear rates calculated using (c) Eyring viscosity and (d) Carreau-Yasuda viscosity. In order to facilitate understanding, the strains are normalized to their values at $t/\lambda = 10^{-2}$. The parameters are set as follows: $\gamma_B = 0.1$, $G = 20\,\text{Pa}$, $G_B = 100\,\text{Pa}$, $\tau_o = 30\,\text{Pa}$, $\eta_o = 70\,\text{Pa}\cdot\text{s}$, $\eta_s = 1\,\text{Pa}\cdot\text{s}$, $\mu = 2.5$, $\nu = 1.5$ and $\lambda = \eta_o/\tau_o$.

Figure 10 shows the creep transition behavior obtained by numerically integrating Eq. (68). Here, $\tau_y$ is the yield stress obtained from the Herschel–Bulkley fit in Fig. 8. As Fig. 8 indicates $\tau_y \neq G_B\gamma_B$ (with $G_B\gamma_B \approx 2\tau_y$ for the present parameter set), one may question whether the yield stress of the Herschel-Bulkley equation remains consistent with the threshold stress for creep transition. However, as shown in Figs. 10(a) and 10(b), creep transition occurs when the applied shear stress exceeds $\tau_y$. If one were to consider $G_B\gamma_B$ as the yield stress for the creep transition, then the transition would be expected to occur near $T_{12}^{\mathrm{app}}/\tau_y \approx 2$. Figures 10(c) and 10(d) provide clear visualizations of the transition. At a low applied shear stress, the shear rate decayed to zero for a long time, whereas for $T_{12}^{\mathrm{app}}/\tau_y > 1$, the shear rate no longer decayed to zero and instead approached a finite constant. Therefore, even in the 3D model, the yield stress obtained from the Herschel-Bulkley fit serves as a reasonable threshold for creep transition.

In summary, Figs. 7-10 show that the 3D constitutive equation satisfies requirements [i]-[iv]. Under start-up shear at a constant shear rate, the shear stress increased with time (or strain) and converged to a steady state (Fig. 7 and requirement [i]). Unlike the 1D model, the 3D constitutive equation exhibited a stress overshoot in the start-up shear, which became more pronounced as the imposed shear rate increased. At low shear rates, the steady state stress as a function of shear rate reflects the Herschel-Bulkley equation (Fig. 8 and requirement [ii]). After the shear stress reached a steady state in the start-up shear, the stress relaxation response at $\dot{\gamma}_{\mathrm{o}} = 0$ approached a non-zero fully relaxed stress for a sufficiently long time (Fig. 9 and requirement [iii]). Both the fully relaxed stress $T_{12}(\infty)$ and normalized fully relaxed stress $T_{12}(\infty)/T_{12}(t_{\mathrm{o}})$ decreased with increasing pre-shearing rate. Finally, in the creep simulations, the response exhibited a clear creep transition as the applied shear stress increased (Fig. 10 and requirement [iv]). Notably, for the parameter set considered here, $G_B \gamma_B$ was approximately twice the yield stress obtained from the Herschel-Bulkley fit $\tau_y$, while the creep transition occured at an applied stress similar to $\tau_y$.

# 5. Conclusion

In this study, we proposed a constitutive equation that describes the characteristic rheological phenomena observed in simple yield-stress fluids such as the Carbopol dispersion. The model reproduced requirements [i]-[iv] introduced in the Introduction (stress overshoot depending on conditions and long-time convergence to a steady state stress, Herschel-Bulkley behavior, non-zero fully relaxed stress, and creep transition).

A non-zero fully relaxed stress of the Carbopol dispersion suggests the possibility of developing a constitutive equation based on viscoelastic solids; however, previous studies pertaining to the development of constitutive equations have been conducted mainly based on viscoelastic fluids. Therefore, we developed a constitutive equation based on the Zener model and introduced an additional linear dashpot connected in parallel to represent solvent-induced viscous dissipation.

To develop a constitutive equation, the strain of the Carbopol dispersion was decomposed into elastic and plastic strains based on its microstructure. Elastic strain represents deformation within individual microgels, whereas plastic strain represents changes in the configuration of aggregated microgels. In the Zener element, the elastic strain corresponds to a series-connected spring, whereas the plastic strain is associated with the deformation of the Kelvin–Voigt element. Additionally, a linear dashpot was connected in parallel to the Zener element to represent the solvent contribution; thus, the total stress was expressed as the sum of the gel stress from the Zener element and the solvent stress. For the 1D model, the additive decomposition of strain was used. To capture the solid-to-liquid transition observed upon yielding, we adopted shear-thinning viscosity models, namely the Eyring and Carreau-Yasuda viscosity models. The plastic strain rate was governed by the flow rule, and the evolution equation of the back stress adhered to the Armstrong-Frederick theory.

Based on analytical and numerical verifications, our 1D constitutive equation satisfies requirements [i]-[iv] in a limited manner. In the start-up shear, the shear stress increased monotonically and approached a steady state value without exhibiting a stress overshoot. At low shear rates, the steady state stress reflected the Herschel-Bulkley equation. During stress relaxation, the stress approached a non-zero fully relaxed state for a long time. The fully relaxed stress was independent of the pre-shearing rate, whereas the normalized fully relaxed stress decreased with increasing pre-shearing rate. During creep, the response exhibited a transition from a viscoelastic solid to a viscoelastic liquid as the applied shear stress increased. The yield stress of the Herschel-Bulkley equation provides a reasonable estimate of the threshold stress.

To apply the proposed constitutive equation to practical processing conditions, the response must be predictable not only under simple shear but also under more general deformation modes. The 1D model was extended to three dimensions by adopting the Kröner-Lee decomposition and satisfying the principle of material frame invariance.

The resulting 3D constitutive equation captures the same key features as the 1D model, including the steady state stress, Herschel-Bulkley behavior, non-zero fully relaxed stress, and creep transition. Moreover, the 3D model exhibited a stress overshoot in the start-up shear and predicted that the fully relaxed stress was a function of the pre-shearing rate. These results suggest that the 3D framework provides a more general description than the 1D formulation. In particular, the stress overshoot in the start-up shear can be predicted from our 3D viscoelastic solid-based constitutive equations without additional assumptions such as isotropic hardening [Dimitriou and McKinley, 2014] or spatially heterogeneous microstructure [Benzi *et al.*, 2021].

The proposed 3D constitutive equation is expected to serve as a foundation for the quantitative analyses of complex flows of yield-stress fluids, such as the Carbopol dispersion as well as electrode slurries used in lithium-ion battery manufacturing. In the latter case, thixotropy should be considered. Furthermore, the equation is expected to facilitate extensions to numerical tools, such as finite-element simulations, under general deformation modes.

# Supplementary Material

This supplementary material contains Figs. S1 and S2. Figure S1 shows the elastic loading term and relaxation term for each condition. Figure S2 shows the gel stress rate for each condition.

# Acknowledgments

We thank Daekwon Jin (Ajou University) for providing the steady-shear data of Carbopol 940 (0.2 wt%).

# Author Declarations

## Conflict of Interest

The authors have no conflicts to disclose.

# Appendix

## Appendix A. Stress-Explicit Approximation to Carreau-Yasuda Model

Beginning from the original Carreau–Yasuda model (Eq. (10)), we obtain the relation between shear stress and shear rate in Eq. (12). Subsequently, Eq. (12) can be rewritten as

$$\tilde{\tau} \equiv \frac{\tau}{\tau_o} = \frac{\tilde{x}}{\left(1+\tilde{x}^{\alpha}\right)^{\beta}} \equiv \phi\left(\tilde{x}\right), \tag{A1}$$

where

$$\lambda \equiv \frac{\eta_o}{\tau_o}; \quad \tilde{x} \equiv \frac{\eta_o \dot{\gamma}}{\tau_o}. \tag{A2}$$

Expressing the inverse of $\phi(x)$ algebraically is not trivial.

The double logarithmic plot of $\tilde{\tau}$ vs. $x$ implies that

$$\frac{d \log \tilde{x}}{d \log \tilde{\tau}} \approx \begin{cases} 1 & \text{for } \tilde{x} \ll 1 \Leftrightarrow \tilde{\tau} \ll 1 \\ 1/(1-n) & \text{for } \tilde{x} \gg 1 \Leftrightarrow \tilde{\tau} \gg 1 \end{cases}. \tag{A3}$$

One of the approximate functions for $\tilde{x} = \phi^{-1}(\tilde{\tau})$ is expressed as

$$\tilde{x} \approx \tilde{\tau}\left(1 + \tilde{\tau}^{\mu}\right)^{\nu}, \tag{A4}$$

where $\mu$ is an adjustable parameter and

$$\frac{1}{1-n} = 1 + \mu\nu > 1. \tag{A5}$$

Equation (A4) can be rewritten adopting the following equations:

$$\dot{\gamma} = \frac{\tau}{\eta_{CY,\tau}(\tau)}, \quad \eta_{CY,\tau}(\tau) = \frac{\eta_o}{\left[1 + \left(\tau/\tau_o\right)^{\mu}\right]^{\nu}}. \tag{A6}$$

Notably, the exponents $\mu$ and $\nu$ do not have to satisfy $0 < \mu < 1$ and $0 < \nu < 1$.

## Appendix B. 3D Model in Simple Shear: Component-wise ODE

In this appendix, we provide the component-wise evolution equations used to numerically integrate Eq. (68) in simple shear. The dimensionless variables defined in Eq. (69) are adopted.

In a simple shear test, the velocity gradient is

$$\tilde{\mathbf{L}} = \frac{d\gamma}{d\tilde{t}} \mathbf{e}_1 \otimes \mathbf{e}_2. \tag{B1}$$

With the initial condition $\mathbf{B}_e(0)=\mathbf{I}$ and $\mathbf{T}_B(0)=\mathbf{O}$, the tensors remain symmetric, and only the components $B_{11}^{(e)}$, $B_{22}^{(e)}$, $B_{33}^{(e)}$, $B_{12}^{(e)}$, $\tilde{T}_{11}^{(B)}$, $\tilde{T}_{22}^{(B)}$, $\tilde{T}_{33}^{(B)}$ and $\tilde{T}_{12}^{(B)}$ need to be integrated. Therefore, the evolution equations are given as follows:

$$\begin{aligned}
\frac{dB_{11}^{(e)}}{d\tilde{t}} &= 2B_{12}^{(e)}\frac{d\gamma}{d\tilde{t}} - 2\left[\tilde{D}_{11}^{(p)}B_{11}^{(e)} + \tilde{D}_{12}^{(p)}B_{12}^{(e)}\right];\\
\frac{dB_{22}^{(e)}}{d\tilde{t}} &= -2\left[\tilde{D}_{12}^{(p)}B_{12}^{(e)} + \tilde{D}_{22}^{(p)}B_{22}^{(e)}\right];\\
\frac{dB_{33}^{(e)}}{d\tilde{t}} &= -2\tilde{D}_{33}^{(p)}B_{33}^{(e)};\\
\frac{dB_{12}^{(e)}}{d\tilde{t}} &= B_{22}^{(e)}\frac{d\gamma}{d\tilde{t}} - \left[\tilde{D}_{11}^{(p)} + \tilde{D}_{22}^{(p)}\right]B_{12}^{(e)} - \tilde{D}_{12}^{(p)}\left[B_{11}^{(e)} + B_{22}^{(e)}\right],
\end{aligned} \tag{B2}$$

$$\begin{aligned}
\frac{d\tilde{T}_{11}^{(B)}}{d\tilde{t}} &= 2\tilde{T}_{12}^{(B)}\frac{d\gamma}{d\tilde{t}} - 2\left[\tilde{D}_{11}^{(p)}\tilde{T}_{11}^{(B)} + \tilde{D}_{12}^{(p)}\tilde{T}_{12}^{(B)}\right] + \tilde{G}_B\tilde{D}_{11}^{(p)} - \frac{\dot{\tilde{e}}_p}{\gamma_B}\tilde{T}_{11}^{(B)};\\
\frac{d\tilde{T}_{22}^{(B)}}{d\tilde{t}} &= -2\left[\tilde{D}_{12}^{(p)}\tilde{T}_{12}^{(B)} + \tilde{D}_{22}^{(p)}\tilde{T}_{22}^{(B)}\right] + \tilde{G}_B\tilde{D}_{22}^{(p)} - \frac{\dot{\tilde{e}}_p}{\gamma_B}\tilde{T}_{22}^{(B)};\\
\frac{d\tilde{T}_{33}^{(B)}}{d\tilde{t}} &= -2\tilde{D}_{33}^{(p)}\tilde{T}_{33}^{(B)} + \tilde{G}_B\tilde{D}_{33}^{(p)} - \frac{\dot{\tilde{e}}_p}{\gamma_B}\tilde{T}_{33}^{(B)};\\
\frac{d\tilde{T}_{12}^{(B)}}{d\tilde{t}} &= \tilde{T}_{22}^{(B)}\frac{d\gamma}{d\tilde{t}} - \left[\tilde{D}_{11}^{(p)} + \tilde{D}_{22}^{(p)}\right]\tilde{T}_{12}^{(B)} - \tilde{D}_{12}^{(p)}\left[\tilde{T}_{11}^{(B)} + \tilde{T}_{22}^{(B)}\right] + \tilde{G}_B\tilde{D}_{12}^{(p)} - \frac{\dot{\tilde{e}}_p}{\gamma_B}\tilde{T}_{12}^{(B)}.
\end{aligned} \tag{B3}$$

# References

Armstrong, P. J. and Frederick, C. O., *A mathematical representation of the multiaxial Bauschinger effect*, (Berkeley Nuclear Laboratories, 1966).

Barnes, H. A., "The yield stress—a review or 'παντα ρει'—everything flows?" *J. Non-Newtonian Fluid Mech.*, **81**, 133-178 (1999).

Benzi, R., Divoux, T., Barentin, C., Manneville, S., Sbragaglia, M., and Toschi, F., "Stress Overshoots in Simple Yield Stress Fluids," *Phys. Rev. Lett.*, **127**, 148003 (2021).

Blatz, P. J., Sharda, S. C., and Tschoegl, N. W., "Strain energy function for rubberlike materials based on a generalized measure of strain," *Trans. Soc. Rheol.*, **18**, 145–161 (1974).

Bonn, D., and Denn, M. M., "Yield stress fluids slowly yield to analysis," *Science*, **324**, 1401-1402 (2009).

Bonn, D., Denn, M. M., Berthier, L., Divoux, T., and Manneville, S., "Yield stress materials in soft condensed matter," *Rev. Mod. Phys.*, **89**, 035005 (2017).

Carreau, P. J., *Rheological equations from molecular network theories*, Ph.D. thesis, University of Wisconsin–Madison (1968).

Cho, K. S., and Kim, S. Y., "Thermodynamic theory of the viscoelasticity and yield of glassy polymers: Internal time theory," *J. Appl. Polym. Sci.*, **89**, 2400–2411 (2003).

Dimitriou, C. J., Ewoldt, R. H., and McKinley, G. H., "Describing and prescribing the constitutive response of yield stress fluids using large amplitude oscillatory shear stress (LAOStress)," *J. Rheol.*, **57**, 27–70 (2013).

Dimitriou, C. J., and McKinley, G. H., "A comprehensive constitutive law for waxy crude oil: a thixotropic yield stress fluid," *Soft Matter*, **10**, 6619-6644 (2014).

Dimitriou, C. J., and McKinley, G. H., "A canonical framework for modeling elasto-viscoplasticity in complex fluids," *J. Non-Newtonian Fluid Mech.*, **265**, 116-132 (2019).

Dinkgreve, M., Denn, M. M., and Bonn, D., "'Everything flows?': elastic effects on startup flows of yield-stress fluids," *Rheol. Acta*, **56**, 189–194 (2017).

Dinkgreve, M., Fazilati, M., Denn, M. M., and Bonn, D., "Carbopol: From a simple to a thixotropic yield stress fluid," *J. Rheol.*, **62**, 773–780 (2018).

Dinkgreve, M., Paredes, J., Denn, M. M., and Bonn, D., "On different ways of measuring 'the' yield stress," *J. Non-Newtonian Fluid Mech.*, **238**, 233–241 (2016).

Divoux, T., Barentin, C., and Manneville, S., "Stress overshoot in a simple yield stress fluid: An extensive study combining rheology and velocimetry," *Soft Matter*, **7**, 9335–9349 (2011).

Eyring, H., "Viscosity, plasticity, and diffusion as examples of absolute reaction rates," *J. Chem. Phys.*, **4**, 283–291 (1936).

Griebler, J. J., Dobo, A. S., Miczuga, E. E., and Rogers, S. A., "Resolving dual processes in complex oscillatory yielding," *Phys. Rev. Lett.*, **135**, 058201 (2025).

Gurtin, M. E., Fried, E., and Anand, L., *The Mechanics and Thermodynamics of Continua*, (Cambridge University Press, 2010).

Haupt, P., *Continuum Mechanics and Theory of Materials*, (Springer, 2002).

Jaworski, Z., Spychaj, T., Story, A., and Story, G., "Carbomer microgels as model yield-stress fluids," *Rev. Chem. Eng.*, **38**, 881–919 (2022).

Kamani, K., Donley, G. J. and Rogers, S. A., "Unification of the rheological physics of yield stress fluids," *Phys. Rev. Lett.*, **126**, 218002 (2021)

Ketz, R. J., Prud'homme, R. K., and Graessley, W. W., "Rheology of concentrated microgel solutions," *Rheol. Acta*, **27**, 531–539 (1988).

Kim, Y., Kim, S., Kim, B. S., Park, J. H., Ahn, K. H., and Park, J. D., "Yielding behavior of concentrated lithium-ion battery anode slurry," *Phys. Fluids*, **34**, 123112 (2022).

Kröner, E., "Allgemeine Kontinuumstheorie der Versetzungen und Eigenspannungen," *Arch. Ration. Mech. Anal.*, **4**, 273–334 (1960).

Lai, W. M., Rubin, D., and Krempl, E., *Introduction to Continuum Mechanics*, (Butterworth-Heinemann, 2010).

Lee, E. H., "Elastic plastic deformation at finite strain," *ASME J. Appl. Mech.*, **36**, 1–6 (1969).

Leonov, A. I., "Nonequilibrium thermodynamics and rheology of viscoelastic polymer media," *Rheol. Acta*, **15**, 85–98 (1976).

Medina-Bañuelos, E., Marín-Santibáñez, B. M., and Pérez-González, J., "Rheo-PIV analysis of the steady torsional parallel-plate flow of a viscoplastic microgel with wall slip," *J. Rheol*, **66**, 31-48 (2022).

Pagani, G., Hofmann, M., Govaert, L. E., Tervoort, T. A., and Vermant, J., "No yield stress required: Stress-activated flow in simple yield-stress fluids," *J. Rheol.*, **68**, 155–170 (2024).

Piau, J. M., "Carbopol gels: Elastoviscoplastic and slippery glasses made of individual swollen sponges Meso- and macroscopic properties, constitutive equations and scaling laws," *J. Non-Newtonian Fluid Mech.*, **144**, 1–29 (2007).

Rosati, L., "A novel approach to the solution of the tensor equation $\mathbf{AX}+\mathbf{XA}=\mathbf{H}$," *Int. J. Solids Struct.*, **37**, 3457–3477 (2000).

Saramito, P., "A new constitutive equation for elastoviscoplastic fluid flows," *J. Non-Newtonian Fluid Mech.*, **145**, 1–14 (2007).

Scheidler, M., “The tensor equation $\mathbf{AX}+\mathbf{XA}=\mathbf{\Phi}(\mathbf{A},\mathbf{H})$, with application to kinematics of continua,” *J. Elasticity*, **36**, 117–153 (1994).

Tervoort, T. A., Klompen, E. T. J., and Govaert, L. E., “A multi-mode approach to finite, three-dimensional, nonlinear viscoelastic behavior of polymer glasses,” *J. Rheol.*, **40**, 779–797 (1996).

Ting, T. C. T., “New expressions for the solution of the matrix equation $\mathbf{A}^T\mathbf{X}+\mathbf{XA}=\mathbf{H}$,” *J. Elasticity*, **45**, 61–72 (1996).

Vinutha, H., Marchand, M., Caggioni, M., Vasisht, V. V., Del Gado, E., and Trappe, V., “Memory of shear flow in soft jammed materials,” *PNAS Nexus*, **3**, 441 (2024).

Yasuda, K. Y., Armstrong, R. C., and Cohen, R. E., “Shear flow properties of concentrated solutions of linear and star branched polystyrenes,” *Rheol. Acta*, **20**, 163–178 (1981).

Zhao, B., Yin, D., Gao, Y., and Ren, J., “Concentration dependence of yield stress, thixotropy and viscoelasticity rheological behavior of lithium-ion battery slurry,” *Ceram. Int.*, **48**, 19073–19080 (2022).

# Supplementary Material

## 1. Decomposition of the Gel Stress Rate into Elastic Loading and Relaxation Terms

To elucidate the mechanism of the stress overshoot, we decompose the gel-stress rate into an elastic loading term and a relaxation term.

For the 1D model,

$$\tau_{\text{gel}} = G\gamma_e, \quad \dot{\tau}_{\text{gel}} = G\left(\dot{\gamma}_{\text{o}} - \dot{\gamma}_p\right), \tag{S1}$$

and therefore

$$E_{1D} \equiv G\dot{\gamma}_{\text{o}}, \quad R_{1D} \equiv G\dot{\gamma}_p, \tag{S2}$$

where $E_{1D}$ and $R_{1D}$ denotes the elastic loading term and the relaxation term for the 1D model, respectively.

For the 3D model under simple shear,

$$\mathbf{T}_{\text{gel}} = G\mathbf{B}_e', \tag{S3}$$

and, since $\left(\mathbf{B}_e'\right)_{12} = B_{12}^{(e)}$,

$$T_{12}^{(\text{gel})} = GB_{12}^{(e)}. \tag{S4}$$

Using the component-wise evolution equation of $\mathbf{B}_e$, the gel stress rate is written as

$$\dot{T}_{12}^{(\text{gel})} = GB_{22}^{(e)}\dot{\gamma}_{\text{o}} - G\left\{\left[D_{11}^{(p)} + D_{22}^{(p)}\right]B_{12}^{(e)} - D_{12}^{(p)}\left[B_{11}^{(e)} + B_{22}^{(e)}\right]\right\}. \tag{S5}$$

Accordingly, we define

$$E_{3D} \equiv GB_{22}^{(e)}\dot{\gamma}_{\text{o}}, \quad R_{3D} \equiv G\left\{\left[D_{11}^{(p)} + D_{22}^{(p)}\right]B_{12}^{(e)} - D_{12}^{(p)}\left[B_{11}^{(e)} + B_{22}^{(e)}\right]\right\}. \tag{S6}$$

For both the 1D and 3D models, we introduce the auxiliary function

$$\text{gel stress rate} \equiv \Phi = E - R. \tag{S7}$$

The condition $\Phi = 0$ is equivalent to $E = R$. Because the solvent stress is constant under imposed start-up shear, the first finite strain zero of $\Phi$ is identical to the first crossing of the elastic loading and relaxation term, and it coincides with the overshoot peak of the total shear stress. To facilitate understanding, the elastic loading term, the relaxation term, and the gel stress rate are non-dimensionalized as follows:

$$\tilde{E} \equiv \frac{\lambda E}{\tau_o}, \quad \tilde{R} \equiv \frac{\lambda R}{\tau_o}, \quad \tilde{\Phi} \equiv \frac{\lambda \Phi}{\tau_o}. \tag{S8}$$

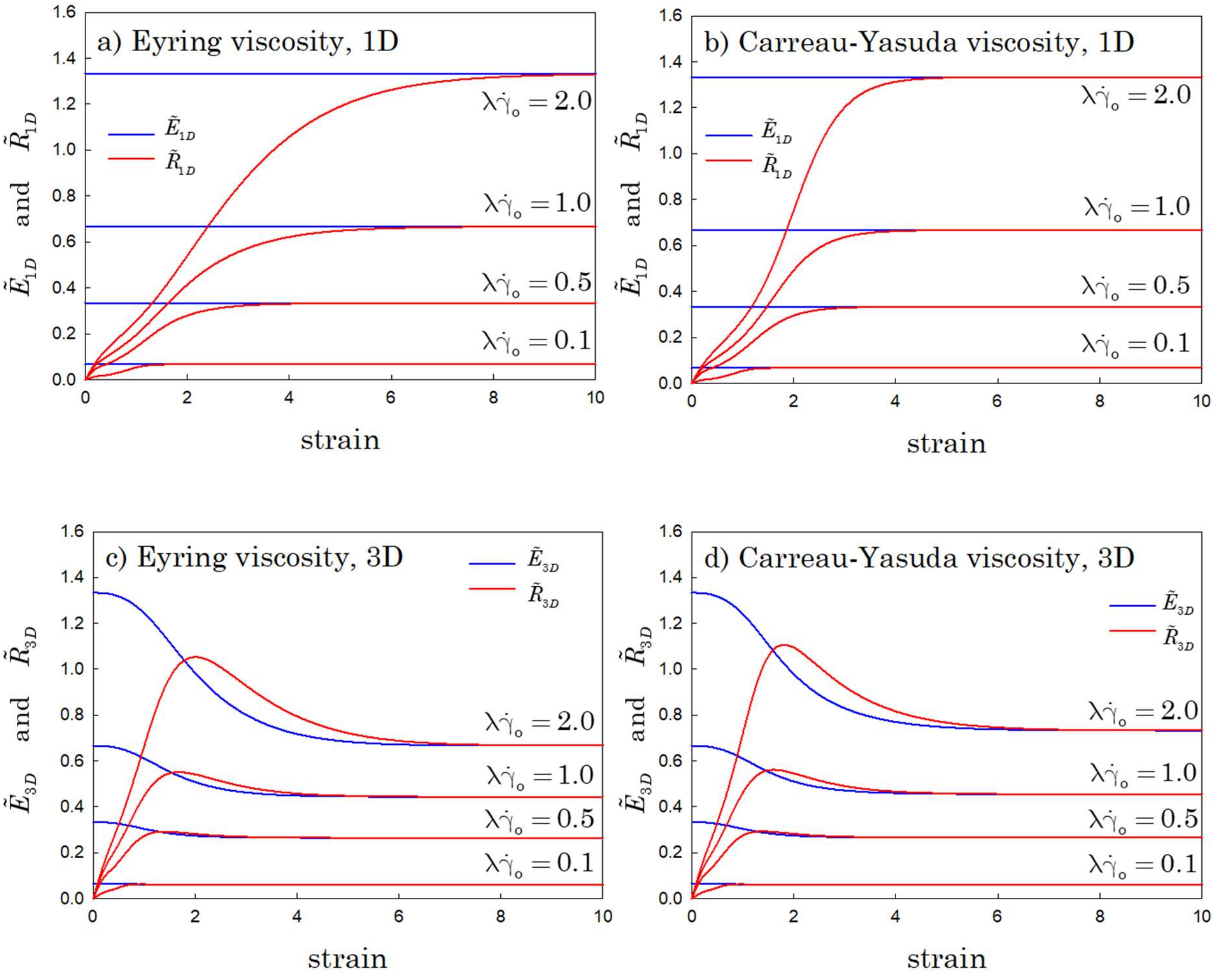

**Fig. S1.** Elastic loading term $E$ and relaxation term $R$ plotted as functions of strain under start-up shear for (a) the 1D model with the Eyring viscosity, (b) the 1D model with the Carreau-Yasuda viscosity, (c) the 3D model with the Eyring viscosity, and (d) the 3D model with the Carreau-Yasuda viscosity, at imposed

shear rate $\lambda\dot{\gamma}_o = 0.1$, 0.5, 1 and 2. The blue lines represent the elastic loading terms and the red lines represent the relaxation terms.

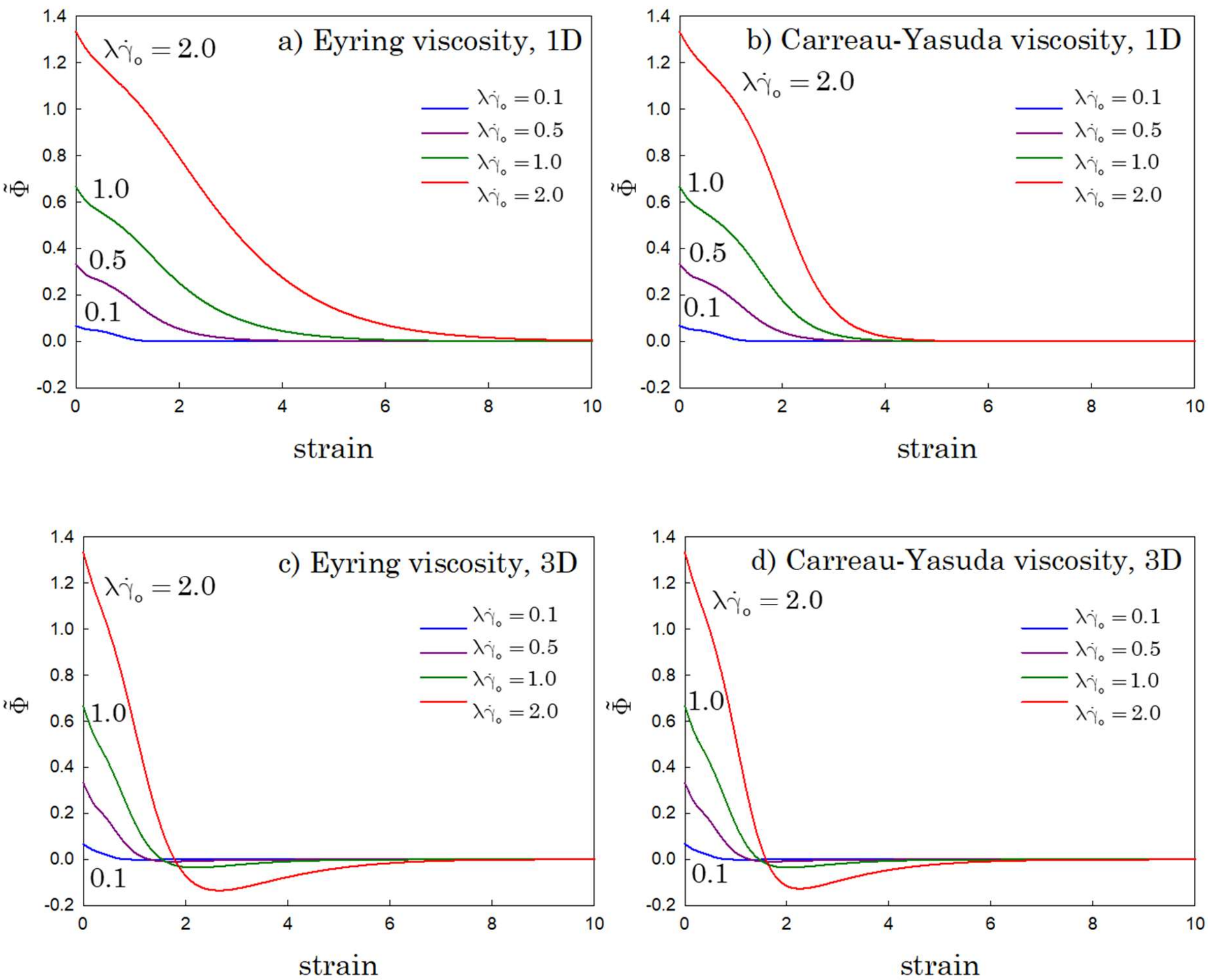

**Fig. S2.** Gel stress rate function $\Phi$ plotted as a function of strain under start-up shear for (a) the 1D model with the Eyring viscosity, (b) the 1D model with the Carreau-Yasuda viscosity, (c) the 3D model with the Eyring viscosity, and (d) the 3D model with the Carreau-Yasuda viscosity, at imposed shear rate $\lambda\dot{\gamma}_o = 0.1$ (blue), 0.5 (purple), 1 (green) and 2 (red).

Figures S1 and S2 show the elastic loading term, the relaxation term, and the corresponding $\Phi$ for the 1D and 3D models with the Eyring and Carreau–Yasuda viscosities at $\lambda\dot{\gamma}_o = 0.1$, 0.5, 1 and 2. For both viscosity models, the 1D model exhibits no finite-strain zero of $\Phi$: the relaxation term approaches the elastic term only asymptotically, and therefore the stress increases monotonically toward the steady state. By contrast, the 3D model exhibits a first finite-strain zero of $\Phi$ at all imposed shear rates considered here. This first zero coincides with the stress peak and directly identifies the onset of stress decrease. Afterward, $\Phi < 0$ over a finite strain interval, showing that the relaxation term temporarily exceeds the elastic loading term. The subsequent convergence of the two terms at larger strain corresponds to the steady-state limit and should not be interpreted as an additional overshoot event.

Taken together, these results provide a direct mechanistic explanation for the difference between the 1D and 3D models. The 1D model lacks a finite-strain regime in which the relaxation term overtakes the elastic loading term, whereas the 3D model exhibits such a regime because of its tensorial coupling. This is why the 1D model does not predict an overshoot, while the 3D model does.